\PassOptionsToPackage{table,dvipsnames}{xcolor}
\documentclass[twocolumn]{aastex631}

\usepackage{xcolor}
\usepackage{lipsum}
\usepackage{xspace}
\usepackage{makecell}

\defcitealias{Tran2022}{DR1}

\usepackage[utf8]{inputenc}
\usepackage[T1]{fontenc}
\usepackage{textcomp}
\usepackage{gensymb}

\rowcolors{2}{gray!15}{white}

\newcommand{\HubbleSNAPnumberlenses}{71\xspace}      %
\newcommand{\HubbleSNAPnumbertotalvisits}{76\xspace} %
\newcommand{\HubbleSNAPnumbertargets}{75\xspace}     %
\newcommand{\HubbleGAPnumberinclBad}{43\xspace}      %
\newcommand{\HubbleGAPnumber}{40\xspace}             %
\newcommand{\NotLensNum}{6\xspace}                   %

\newcommand{\HubbleSnapSomeSpec}{50\xspace}          %

\newcommand{\TotalGoodHSTimagesLenses}{111\xspace}  %

\newcommand{\HuangGoodImages}{51\xspace}

\newcommand{\DRoneTotalLensSystems}{68\xspace}            %
\newcommand{\TotalLensSystems}{138\xspace}            %

\newcommand{\TotalNumMeasuredRedshifts}{204\xspace}   %
\newcommand{\NumSystemsBothSrcDe}{101\xspace}
\newcommand{\NumSystemsJustSrc}{32\xspace}
\newcommand{\NumSystemsJustDe}{5\xspace}
\newcommand{\XhsooterTotalRedshifts}{62\xspace}
\newcommand{\KCWITotalRedshifts}{14\xspace}
\newcommand{\ESITotalRedshifts}{51\xspace}
\newcommand{\NIRESTotalRedshifts}{67\xspace}
\newcommand{\MOSFIRETotalRedshifts}{9\xspace}
\newcommand{\DEIMOSTotalRedshifts}{1\xspace}
\newcommand{\TotalRedshiftsLowQuality}{12\xspace}
\newcommand{\newredshiftsDRonevsDRtwo}{106\xspace}
\newcommand{\numzSRCzDEandHST}{69\xspace}

\newcommand{\NumGalaxyScaleLenses}{36\xspace} %

\newcommand{\NumGalaxyScaleInBothHSTsnapPrograms}{125\xspace}
\newcommand{\NumDoubleSourceGalaxyScaleInBothHSTsnapPrograms}{6\xspace}

\newcommand{\SuccessRate}{96\xspace}  %

\newcommand{\doublesourcenum}{6\xspace}

\newcommand{\ASTROthreeD}{The ARC Centre of Excellence for All Sky Astrophysics in 3 Dimensions (ASTRO 3D), Australia}
\newcommand{\swinburne}{Centre for Astrophysics and Supercomputing, Swinburne University of Technology, PO Box 218, Hawthorn, VIC 3122, Australia}
\newcommand{\harvard}{Center for Astrophysics, Harvard \& Smithsonian, Cambridge, MA 02138, USA}
\newcommand{\usyd}{Sydney Institute for Astronomy, School of Physics, A28, The University of Sydney, NSW 2006, Australia}
\newcommand{\ucdavis}{Department of Physics and Astronomy, University of California, Davis, 1 Shields Avenue, Davis, CA 95616, USA}

\begin{document}

\title{The AGEL Survey Data Release 2: A Gravitational Lens Sample for Galaxy Evolution and Cosmology}

\correspondingauthor{Tania M. Barone}
\email{tbarone@swin.edu.au}
\author[0000-0002-2784-564X]{Tania M. Barone}
\affiliation{\swinburne}
\affiliation{\ASTROthreeD}

\author[0000-0002-2645-679X]{Keerthi Vasan G.C.}
\affiliation{\ucdavis}
\affiliation{The Observatories of the Carnegie Institution for Science, 813 Santa Barbara Street, Pasadena, CA 91101, USA}

\author[0000-0001-9208-2143]{Kim-Vy Tran}
\affiliation{\harvard}
\affiliation{\ASTROthreeD}

\author[0000-0003-1362-9302]{Glenn G. Kacprzak}
\affiliation{\swinburne}
\affiliation{\ASTROthreeD}

\author[0000-0002-3254-9044]{Karl Glazebrook}
\affiliation{\swinburne}
\affiliation{\ASTROthreeD}

\author[0000-0001-5860-3419]{Tucker Jones}
\affiliation{\ucdavis}

\author[0009-0008-6114-1401]{Duncan J. Bowden}
\affiliation{\harvard}
\affiliation{School of Physics \& Astronomy, University of Southampton, Southampton SO17 1BJ, UK}

\author[0009-0000-4328-200X]{Faith Dalessandro}
\affiliation{\ucdavis}

\author[0000-0003-0234-6585]{Nandini Sahu}
\affiliation{University of New South Wales, Sydney, NSW 2052, Australia}
\affiliation{\ASTROthreeD}

\author[0000-0003-0516-3485]{Hannah Skobe}
\affiliation{Department of Physics, McWilliams Center for Cosmology and Astrophysics, Carnegie Mellon University, 5000 Forbes Avenue, Pittsburgh, PA 15213, USA}

\author[0000-0002-7278-9528]{Rebecca J. Allen}
\affiliation{\swinburne}

\author[0009-0000-6139-3280]{A. Makai Baker}
\affiliation{School of Physics and Astronomy, Monash University, Clayton VIC 3800, Australia}

\author[0009-0003-3198-7151]{Daniel J. Ballard}
\affiliation{\usyd}

\author[0000-0003-4520-5395]{Yuguang Chen}
\affiliation{\ucdavis}
\affiliation{Department of Physics, The Chinese University of Hong Kong, Shatin, N.T., Hong Kong SAR, China}

\author[0000-0001-5564-3140]{Thomas E. Collett}
\affiliation{Institute of Cosmology and Gravitation, University of Portsmouth, Burnaby Rd, Portsmouth PO1 3FX, UK}

\author[0000-0002-2012-4612]{Giovanni Ferrami}
\affiliation{School of Physics, University of Melbourne, Parkville, VIC 3010, Australia}
\affiliation{\ASTROthreeD}

\author[0000-0001-7282-3864]{Jimena Gonz\'alez}
\affiliation{Physics Department, University of Wisconsin-Madison. 2320 Chamberlin Hall. Madison, WI 53706-1390, USA}

\author[0009-0008-5372-1318]{William Gottemoller}
\affiliation{\harvard}

\author[0000-0001-9414-6382]{Anishya Harshan}
\affiliation{University of Ljubljana, Department of Mathematics and Physics, Jadranska ulica 19, SI-1000 Ljubljana, Slovenia}

\author[0000-0001-8156-0330]{Xiaosheng Huang}
\affiliation{Department of Physics \& Astronomy, University of San Francisco, 2130 Fulton Street, San Francisco, CA 94117-1080, USA}
\affiliation{4 Physics Division, Lawrence Berkeley National Laboratory, 1 Cyclotron Road, Berkeley, CA 94720, USA}

\author[0009-0006-4812-2033]{Leena Iwamoto}
\affiliation{\harvard}

\author[0000-0003-4239-4055]{Colin Jacobs}
\affiliation{\swinburne}
\affiliation{\ASTROthreeD}

\author[0000-0001-6089-0365]{Tesla E. Jeltema}
\affiliation{University of California, Santa Cruz, Santa Cruz, CA 95064, USA}
\affiliation{Santa Cruz Institute for Particle Physics, Santa Cruz, CA 95064, USA}

\author[0009-0008-3531-900X]{Kaustubh Rajesh Gupta}
\affiliation{\swinburne}

\author[0000-0003-3081-9319]{Geraint F. Lewis}
\affiliation{\usyd}

\author[0000-0003-0389-0902]{Sebastian Lopez}
\affiliation{Departamento de Astronom\'ia, Universidad de Chile, Casilla 36-D, Santiago, Chile}

\author[0000-0003-2804-0648]{Themiya Nanayakkara}
\affiliation{\swinburne}
\affiliation{\ASTROthreeD}

\author[0000-0003-2377-8352]{Nikole M. Nielsen}
\affiliation{Homer L. Dodge Department of Physics and Astronomy, The University of Oklahoma, 440 W. Brooks St., Norman, OK 73019, USA}
\affiliation{\swinburne}
\affiliation{\ASTROthreeD}

\author[0000-0003-4083-1530]{Jackson O'Donnell}
\affiliation{University of California, Santa Cruz, Santa Cruz, CA 95064, USA}
\affiliation{Santa Cruz Institute for Particle Physics, Santa Cruz, CA 95064, USA}

\author[0009-0006-0299-0265]{Huimin Qu}
\affiliation{\usyd}

\author[0009-0007-0184-8176]{Sunny Rhoades}
\affiliation{\ucdavis}

\author[0000-0002-5558-888X]{Anowar Shajib}
\affiliation{Department of Astronomy \& Astrophysics, University of Chicago, Chicago, IL 60637, USA}
\affiliation{Kavli Institute for Cosmological Physics, University of Chicago, Chicago, IL 60637, USA}

\author[0000-0002-1576-2505]{Sarah M. Sweet}
\affiliation{School of Mathematics and Physics, University of Queensland, Brisbane, QLD 4072, Australia}
\affiliation{\ASTROthreeD}

\author[0000-0002-1883-4252]{Nicolas Tejos}
\affiliation{Instituto de F\'isica, Pontificia Universidad Cat\'olica de Valpara\'iso, Casilla 4059, Valpara\'iso, Chile}

\begin{abstract}
The ASTRO 3D Galaxy Evolution with Lenses (AGEL) Survey is an ongoing effort to spectroscopically confirm a diverse sample of gravitational lenses with high spatial resolution imaging, to facilitate a broad range of science outcomes. The AGEL systems span single galaxy-scale deflectors to groups and clusters, and include rare targets such as galaxy-scale lenses with multiple sources, lensed quiescent galaxies, and Einstein rings. We build on the 77 systems presented in Tran et al. 2022 (AGEL data release 1) to present a total \TotalLensSystems lenses, and high resolution F140W and F200LP Hubble Space Telescope images for \HubbleSNAPnumberlenses lenses from a completed HST SNAP program. Lens candidates were originally identified by convolutional neural networks in the DES and DECaLS imaging fields, and of the targets with follow-up spectroscopy we find a high (\SuccessRate\%) success rate. Compared with other spectroscopic lens samples, AGEL lenses tend to have both higher redshift deflectors and sources. We briefly discuss the common causes of false-positive candidates, and strategies for mitigating false-positives in next generation lens searches. Lastly, we present \doublesourcenum galaxy-scale double-source plane lenses useful for cosmological analyses. With next-generation telescopes and surveys such as Euclid, Vera Rubin's Legacy Survey of Space and Time, Keck Observatory's KAPA program, and 4MOST's  4SLSLS surveys on the horizon, the AGEL survey represents a pathfinder for refining automated candidate search methods and identifying and triaging candidates for followup based on scientific potential.

\end{abstract}


\section{Introduction} \label{sec:intro}

Strong gravitational lenses are powerful tools to study the most elusive aspects of our Universe. The lensing configuration and magnification of the background source depends upon (i) the total mass distribution of the foreground lens (the `deflector') which range from individual galaxies to clusters, and (ii) the geometry of the universe in the angular diameter distances between us, the deflector, and the source. As a result strong gravitational lenses are used to explore a broad range of astrophysical phenomena in both the lensed source and the foreground deflector, as well as other targets along the line of sight and the cosmological nature of our Universe.

With the deflectors acting as cosmic-scale magnifying glasses, the bright and highly magnified source galaxies offer a rare view of early galaxy formation and evolution at high-redshift \citep[$ z > 1$; e.g. ][]{Zheng2012,McLeod2015,Zhuang2022,Vasan2023} yielding high spatial resolutions (sub-kpc scales) out of reach for even the most powerful telescopes \citep[e.g.][]{Ritondale2019a}. Source magnifications ($\mu$) of over an order of magnitude are commonly reached by galaxy-scale lenses \citep{Shu2016b}, while lensing by clusters can reach magnifications of $\mu = 10-100$ \citep{Yuan2012,Coe2013,Atek2015,Ebeling2018,Florian2021,Sharon2022a,Bergamini2023,Price2024}.

Lensing is sensitive to all mass along the line of sight, so strong lenses provide a way to probe the distribution of both dark and baryonic matter in the deflector. The total mass density profile of massive quiescent galaxies reflects their formation history, with the slope of the profile evolving with redshift due to the two-phases of their formation: mass growth driven by in-situ star formation at high redshift, followed by mass growth through mergers \citep{Oser2010}. Therefore the evolution with redshift of quiescent galaxy total mass density profiles is a key probe of galaxy evolution \citep[e.g.][]{Ruff2011,Bolton2012,Sonnenfeld2013b,Remus2017,Wang2019,Sahu2024}.

In addition to the dark and baryonic matter within the galaxy, the faint diffuse gas surrounding line of sight galaxies in their circumgalactic mediums leaves absorption signatures in the spectrum of the brightly lensed background sources \citep[e.g.][]{Zahedy2016,Lopez2020,Mortensen2021}. With precise lens modelling of high resolution imaging, the spatial distribution of diffuse circumgalactic gas can also be measured from its gravitational effect \citep{Barone2024}.

From precision lens models we can also search for and constrain the scale of dark matter substructure. Dark matter subhalos along the line of sight lead to subtle perturbations in the resulting lens configuration, which can be detected with precise lens modelling on high spatial resolution imaging. Both warm and cold dark matter models predict that galaxies should contain thousands of dark matter subhalos, with $\Lambda$CDM predicting a higher abundance of low-mass halos than warmer models \citep{Bode_Ostriker_Turok2001}. Therefore detections at low halos masses will allow us to distinguish between theoretical dark matter models \citep[e.g.][]{Vegetti2012, Nightingale2022, Lagattuta2023, Bayer2023, Hughes2024}.

Lastly, gravitational lenses can also be used to infer the geometry and composition of the Universe at large scales. The time delay between multiply lensed images of variable sources such as supernovae \citep{Refsdal1964,Goobar2002,More2017,Bayer2021} and quasars \citep[e.g.][]{Suyu2010,Wong2020} is sensitive to the Hubble constant $H_0$, providing an additional measurement of $H_0$ independent from the distance ladder and the cosmic microwave background (CMB). Other cosmological parameters including the dark energy equation of state $w$ can be measured from lenses with multiple sources \citep{Collett_Auger2014,Caminha2022}. This method, termed gravitational lens cosmography \citep{Blandford_Narayan1992,Treu2010_annual_review}, also provides constraints on the total matter density of the Universe, $\Omega_m$, and the curvature, $\Omega_k$, that are nearly orthogonal to those from the CMB and baryon acoustic oscillations \citep{Eisenstein2005,Percival2010}, proving these methods to be highly complimentary.

\subsection{Lens surveys}

The difficulty in many of these science areas is the limited samples of high quality spectroscopically confirmed lenses. Bright (R$_{\rm{AB}}  \leq 22$ mag) systems are rare \citep[$\lesssim 0.1$ per square degree;][]{Jacobs2019a,Jacobs2019b, X_Huang2020}, and therefore discovering new lenses requires wide area imaging over 100s of square degrees. Confirming strong gravitational lenses often requires both high resolution subarcsecond imaging and spectroscopy. Therefore, although simulations predict order $\sim 10^3$ more galaxy-galaxy lenses should be detectable in currently available wide-field imaging surveys \citep{Collett2015}, the number of spectroscopically confirmed lenses is far smaller (few 100s).

The first generation of dedicated lens searches in wide-field imaging surveys was made possible by SDSS \citep[the Sloan Digital Sky Survey;][]{York2000}. SLACS \citep[Sloan Lens ACS Survey;][]{Bolton2008} and BELLS \citep[BOSS Emission-Line Lens Survey;][]{Brownstein2012} confirmed first 70 then another 36 lenses in SDSS, identified by searching for high redshift emission lines within the fibre spectroscopy of a lower redshift target. While this method had the advantage of immediate spectroscopic redshift confirmation, only systems with small Einstein radii (smaller than the diameter of the fiber, so $\rm{R_E} <1.5 \arcsec$ for SLACS and $\rm{R_E} < 1 \arcsec$ for BELLS) were captured and galaxy-scale lenses with larger Einstein radii and group or cluster lenses were missed.

\begin{figure*}
    \centering
    \includegraphics[width=\textwidth]{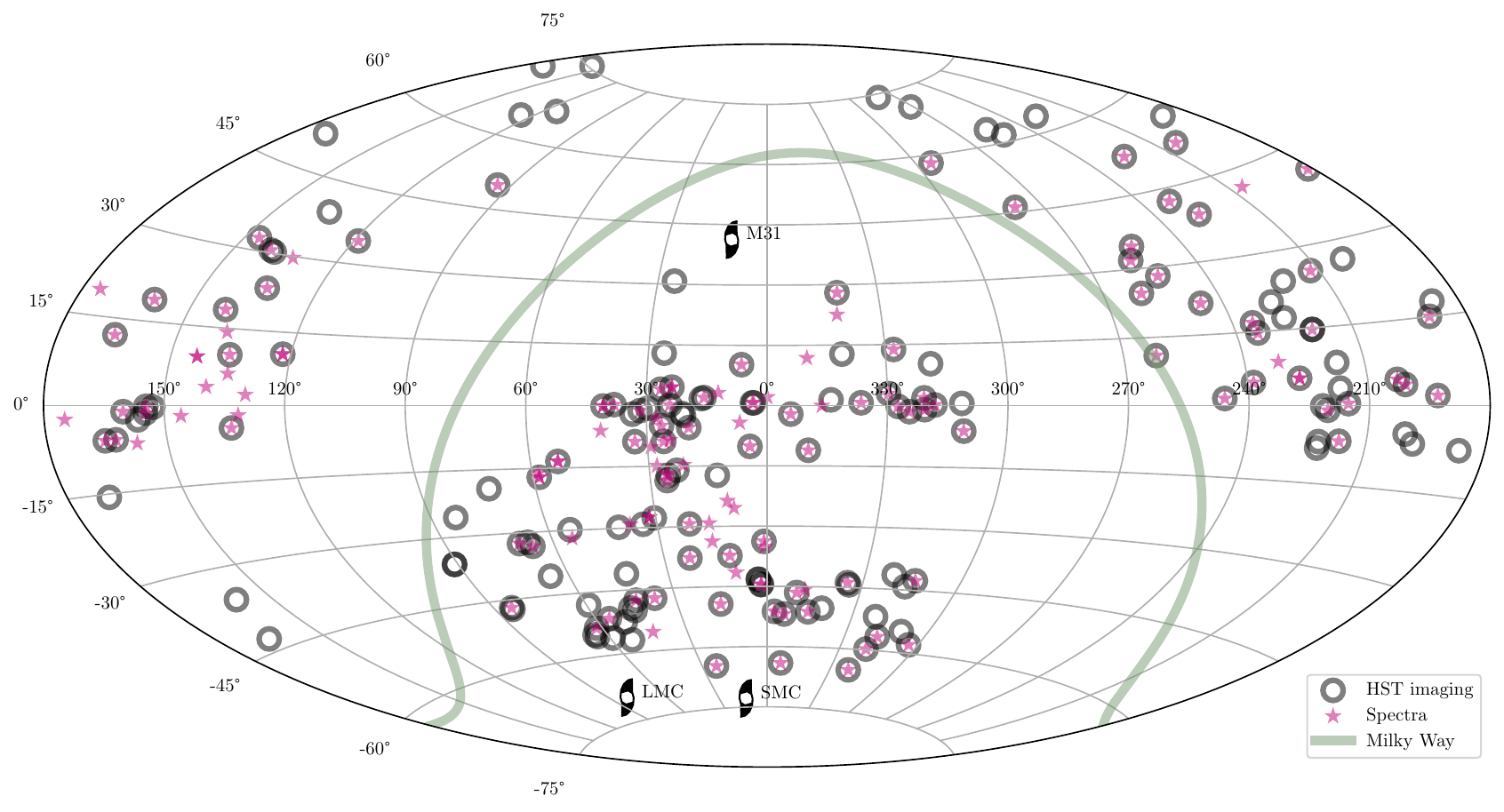}
    \caption{Sky location of targets with spectroscopic follow-up (magenta stars) and HST imaging (black circles) from programs 16773 (PI Glazebrook), 17307 (PI Tran) and 15867 (PI Huang). The plane of the Milky Way is shown in green. The Large and Small Magellanic Clouds and M31 are shown by black swirl symbols.}
    \label{fig:sky_map}
\end{figure*}

Imaging has proven to be a powerful and complementary method of identifying lens candidates. Early searches for cluster-scale lenses relied on visual identification, such as SGAS \citep[the Sloan Giant Arc Survey;][]{Hennawi2008,Sharon2020}. However finding galaxy and group-scale lenses requires search methods that can be easily scaled to large datasets. As a result, teams began using dedicated algorithms to search for blue arcs around massive red foreground galaxies in large imaging surveys. The CASSOWARY survey \citep[the CAmbridge Sloan Survey Of Wide ARcs in the skY;][]{Stark2013_CASSOWARY} used SDSS imaging to search for blue sources within a specified separation from luminous red galaxies \citep{Belokurov2009}. Their search yielded 45 objects, including now well known lenses such as the Cosmic Horseshoe \citep{Belokurov2007}. The SL2S team \citep[the Strong Lensing Legacy Survey;][]{Cabanac2007} developed this technique further and used separate algorithms optimised to search for giant arcs around groups and clusters (\citealt{More2012}, using \textsc{ArcFinder}; \citealt{Alard2006}) and for galaxy-scale lens (\citealt{Gavazzi2012}, using \textsc{RingFinder}; \citealt{Gavazzi2014}) in the CFHT (Canada-France-Hawaii Telescope) Survey. Although these methods led to many successful identifications, due to their proscriptive nature, they can only find systems that fit the tight search parameters. As a result, systems with red sources (due either to dust or their quiescent nature) are disfavoured, as well as systems with small, truncated arcs.

As wide-field area surveys have improved (in depth and breadth) and the demand for larger and more diverse lens samples increased, lens candidate searches are now relying on sophisticated machine learning methods, specifically convolutional neural networks \citep[CNNs;][]{LeCun1989}, to sort through the terabytes of images to find the 1 in $\sim$100,000 galaxies that is strongly lensed \citep{Jacobs2019b}. CNNs have been successfully applied to the CFHTLS \citep{Jacobs2017}, KiDS \citep{Petrillo2017,Petrillo2019,Li2020}, DES \citep{Jacobs2019a,Jacobs2019b}, PanSTARRS \citep{Canameras2020} and DECaLS \citep{X_Huang2020,X_Huang2021} imaging surveys, returning thousands of new lens candidates. The CNN approach has proven to be highly successful in producing high-fidelity candidate samples that were missed by previous searches despite being comparably bright \citep{Tran2022}.

Over the next decade approximately $\sim 10^6$ galaxy-galaxy strong lenses will be discovered by cosmological surveys such as Euclid and the Legacy Survey of Space and Time \citep{Collett2015}, a two orders of magnitude increase in the number of lens candidates currently known and a three orders of magnitude increase over the hundreds of systems that are currently spectroscopically confirmed \citep{Bolton2008,Sonnenfeld2013a,Bolton2012,Shu2016a}. In this coming era, it will become even more important to hone and refine lens search methods.

While coming surveys will provide a plethora of lens images, for many science cases spectroscopic confirmation and redshift measurement is still a crucial requirement. The future 4MOST (4 meter Multi-Object Spectroscopic Telescope) facility will obtain spectra for $\sim 10^4$ of these lenses via 4SLSLS \citep[The 4MOST Strong Lensing Spectroscopic Legacy Survey;][]{Collett2023}, however that represents only 1\% of the total candidate sample. Furthermore, similarly for SLACS and BELLS, only targets with Einstein radii smaller than the fiber (radius $=0.725\arcsec$) will the spectrum capture both the source and deflector light. Importantly, because lenses are so sparse across the sky, 4SLSLS will be sharing the focal plane of 4MOST with other surveys and as a result target choice is limited. Therefore dedicated spectroscopy of high-value targets will still be required.

\subsection{The ASTRO3D Galaxy Evolution with Lenses Survey}

The goal of the ASTRO 3D Galaxy Evolution with Lenses (AGEL) Survey \citep{Tran2022} is to spectroscopically confirm a diverse range of candidate lenses to enable a diverse range of science. %
In the first AGEL data release \citep{Tran2022} we published a catalogue of 104 new redshifts in a total \DRoneTotalLensSystems systems. Combined with literature redshifts, we presented 53 lens systems with deflector and source redshifts, and 15 systems with either a source or deflector spectroscopic redshift. In this second release we present an additional \newredshiftsDRonevsDRtwo new redshifts in a total \TotalLensSystems systems. Combined with literature redshifts, \NumSystemsBothSrcDe targets have both source and deflector redshifts, \NumSystemsJustSrc have just a source redshift and \NumSystemsJustDe systems have only a deflector redshift.

In addition to the expanded spectroscopic catalogue we present \TotalGoodHSTimagesLenses high resolution HST images as part of 2 programs (16773 PI Glazebrook, 17307 PI Tran), as well as an additional \HuangGoodImages images from literature program 15867 \citep[PI Huang;][]{Huang2025}.

Targets span the full range of right ascensions and have declinations ranging between $-65 \degree < \delta < 75 \degree $; the sky distribution of targets is shown in Figure \ref{fig:sky_map}. Candidate lenses were primarily chosen from \cite{Jacobs2019a,Jacobs2019b} but also include targets from \cite{X_Huang2020,X_Huang2021} and \cite{Diehl2017}. Systems are named by their right ascension and declination with a letter A, B, C, differentiating between multiple sources.

The paper is structured as follows. In Section \ref{sec:imaging} we present the \TotalGoodHSTimagesLenses HST images and \HuangGoodImages literature images. We then summarise the observations used for measuring redshifts including the various instrument setups and present the redshift catalogue in Tables \ref{tbl:redshift_summary} and \ref{tbl:redshift_summary_extended}. In Section \ref{sec:sample_characteristics} we discuss the sample characteristics: firstly discussing false-positives lens candidates in \ref{sec:imposters}, followed by a comparison of the AGEL redshift distribution to other lensing surveys and expectations from an analytical model in  \ref{sec:comparison_other_surveys}. Lastly in Section \ref{sec:double_source_lenses} we present galaxy-scale lenses we have found that have multiple source planes. All magnitudes are AB magnitudes. We assume a flat $\Lambda$ cold dark matter ($\Lambda$CDM) Universe with $\Omega_\Lambda= 0.7$, $\Omega_M= 0.3$, and $H_0 =70 $kms$^{-1}$Mpc$^{-1}$. Redshifts are quoted with respect to the solar system's barycenter.

\section{High Resolution HST Imaging}\label{sec:imaging}

\begin{figure}
    \centering
    \includegraphics[width=\linewidth]{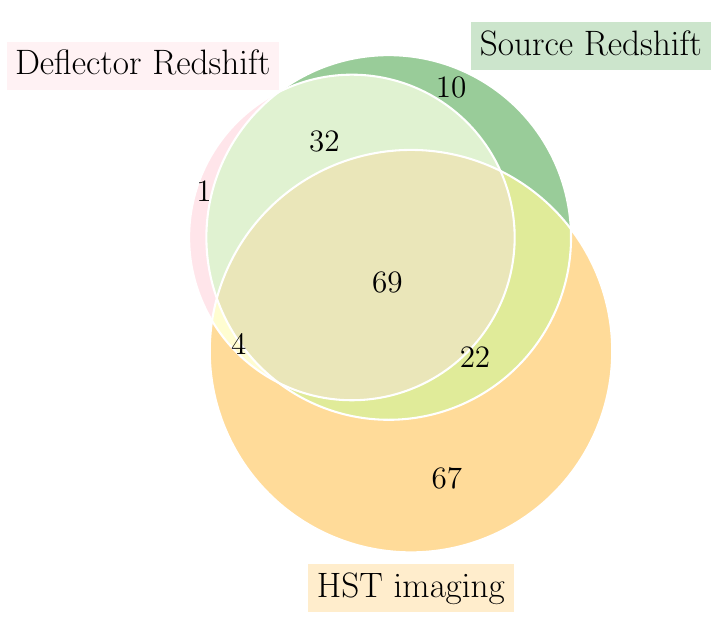}
    \caption{Venn diagram of targets with source redshifts, deflector redshifts, and HST imaging from completed AGEL large program 16773 (PI: Glazebrook), ongoing AGEL large program 17307 (PI: Tran) and archival SNAP program targeting gravitational lenses 15867 (PI: Huang).}
    \label{fig:venn_diagram}
\end{figure}

\begin{table*}
\resizebox{\textwidth}{!}{
\begin{tabular}{ccl}
\hline
AGEL Name & HST object name & Comment \\
(1) & (2) & (3) \\
\hline
\multicolumn{3}{c}{Program 16773, PI Glazebrook}\\
\hline
AGEL211515+101153A & DCLS2115+1011 & Not a lens:  ring galaxy \\
AGEL224400+124540A & DCLS2244+1245 & Not a lens: blue blob adjacent to red star \\
AGEL131330-064211A & DCLS1313-0642 & Not a lens: blue blobs adjacent to red galaxy \\
AGEL132557+263659A & DCLS1325+2636 & Not a lens: blue blobs adjacent to red galaxy \\
AGEL105100-055628A & DLS432021848 & Guiding issue: no usable UVIS data \\
AGEL140839+253104A & DCLS1408+2531 & Guiding issue: lost UVIS data from first visit but was revisited \\
AGEL001030-431515A & DESJ0010-4315 & Guiding issue: has only a single useable IR exposure \\
AGEL022709-471856A & DESJ0227-4718 & Used original observing stratergy \\
AGEL025029-410418A & DESJ0250-4104 & Used original observing stratergy \\
AGEL034131-513045A & DESJ0341-5130 & Used original observing stratergy \\
AGEL035346-170639A & DCLS0353-1706 & Used original observing stratergy \\
AGEL032904-565658A & DESJ0329-5656 & Used original observing stratergy \\
AGEL013355-643413A & DESJ0133-6434 & Used original observing stratergy \\

\hline
\multicolumn{3}{c}{Program 17307, PI Tran}\\
\hline
AGEL180556+705719A & DESI-271.4849+70.9553 & No usable data \\
AGEL033631-295144A & DES0337-295 & No usable data \\
AGEL014556+040229A & DESJ0145+0402 & incorrect sky coordinates target not observed \\

\hline
\multicolumn{3}{c}{Program 15867, PI Huang}\\
\hline
AGEL103255+751854A & DESI-158.2306+75.3149 & No usable data \\
AGEL142822+031800A & DESI-217.0936+03.3000 & No usable data \\
AGEL085331+232155A & DESI-133.3800+23.3652 & Some usable data \\

\hline
\end{tabular}
}\caption{Details of targets with data issues from HST programs 16773, 17307, and 15867. (1) AGEL object name, (2) object name used in the HST observations, (3) target details.}\label{tbl:hst_snap_comments}
\end{table*}

\begin{figure*}
    \centering
    \includegraphics[width=\textwidth]{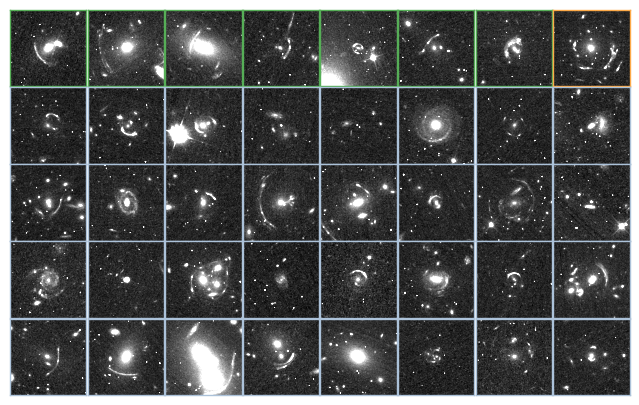}
   \caption{The \HubbleGAPnumber lenses imaged by HST GAP program 17307 (PI Tran). Targets bordered in green have confirmed deflector and source redshifts in the AGEL catalogue, targets bordered in orange have either a source or deflector redshift, while a grey border indicates no redshifts in AGEL. All images are $20\arcsec \times 20\arcsec$ with north oriented up and east left. Object names are listed in Table \ref{tbl:GAP_hist_fig}.}
    \label{FIG:GAP_HST_FIG}
\end{figure*}

As part of the survey we have obtained two HST imaging programs: a completed SNAP (ID: 16773, PI: Glazebrook, cycles 29-30, see Figure \ref{FIG:HST_SNAP_MEGAFIG}) awarded 150 orbits with WFC3 and an ongoing GAP program (ID: 17307, PI: Tran, cycles 32-37, see Figure \ref{FIG:GAP_HST_FIG}) awarded 500 orbits with ACS. The goal of these programs is to build a legacy library of bright strong gravitational lenses that span galaxy to cluster-scale halos. Given the resolution of HST's WFC3 and ACS (point spread functions of $0.124-0.156 \arcsec$, $0.083 - 0.089 \arcsec$ and $0.1-0.14$ for WFC3/IR, WFC3/UVIS, and ACS/WCF respectively), HST has a distinct advantage for determining accurate lens models over ground-based surveys such as DES, HSC, and the upcoming LSST, as well as space-based imaging surveys with Euclid and Roman (both $0.1\arcsec/$pixel). If we consider these two programs as well as imaging from archival SNAP program 15867 (PI: Huang) a total of \numzSRCzDEandHST targets have the full combination of source redshift, deflector redshift, and HST imaging (see Figure \ref{fig:venn_diagram}).

One of the primary motivations for the high-resolution follow-up imaging is to enable reliable lens models, which are necessary to access the full scope of science from lenses. The precision on the magnification inferred from lens models on ground-based imaging can be improved 10-fold with HST imaging (cf. $\mu = 77^{+43}_{-26}$ from \cite{Sukay2022} to $\mu = 14.6^{+0.55}_{-1.07}$ from \cite{Zhuang2022} for the same system). Precise measurements of the lensing magnification are needed to constrain source properties such as the stellar mass and star formation rate. Furthermore, high resolution imaging helps confirm the lensing nature of the target and the multiplicity and morphology of the lensed source. Thus far, AGEL targets ranging from galaxy to group and cluster scale lenses have been modelled thanks to this new HST imaging \citep{Zhuang2022,Sahu2024,Sheu2024}. While all SNAP and GAP data has zero propriety period and so the raw data is all publicly available, we  provide reduced files as well as RGB images on the AGEL Survey website\footnote{\href{https://sites.google.com/view/agelsurvey/science/hubble-images}{https://sites.google.com/view/agelsurvey/science/hubble-images}}.

While not all targets in the AGEL spectroscopic catalogue have HST imaging, all targets have \textit{grz} imaging from the Dark Energy Survey \citep[DES;][]{Abbott2018} and the The Dark Energy Camera Legacy Survey \citep[DECaLS;][]{Dey2019} by virtue of the selection method (see Section \ref{sec:spectroscopic_data}). The DES and DECaLS survey span 9000 deg$^2$ and reaches 5$\sigma$ depths of $g=24.0$, $r=23.4$ and $z=22.5$ mags.

\subsection{WFC3 SNAP Program 16773, PI Glazebrook}
\begin{figure*}
    \centering
    \includegraphics[width=\textwidth]{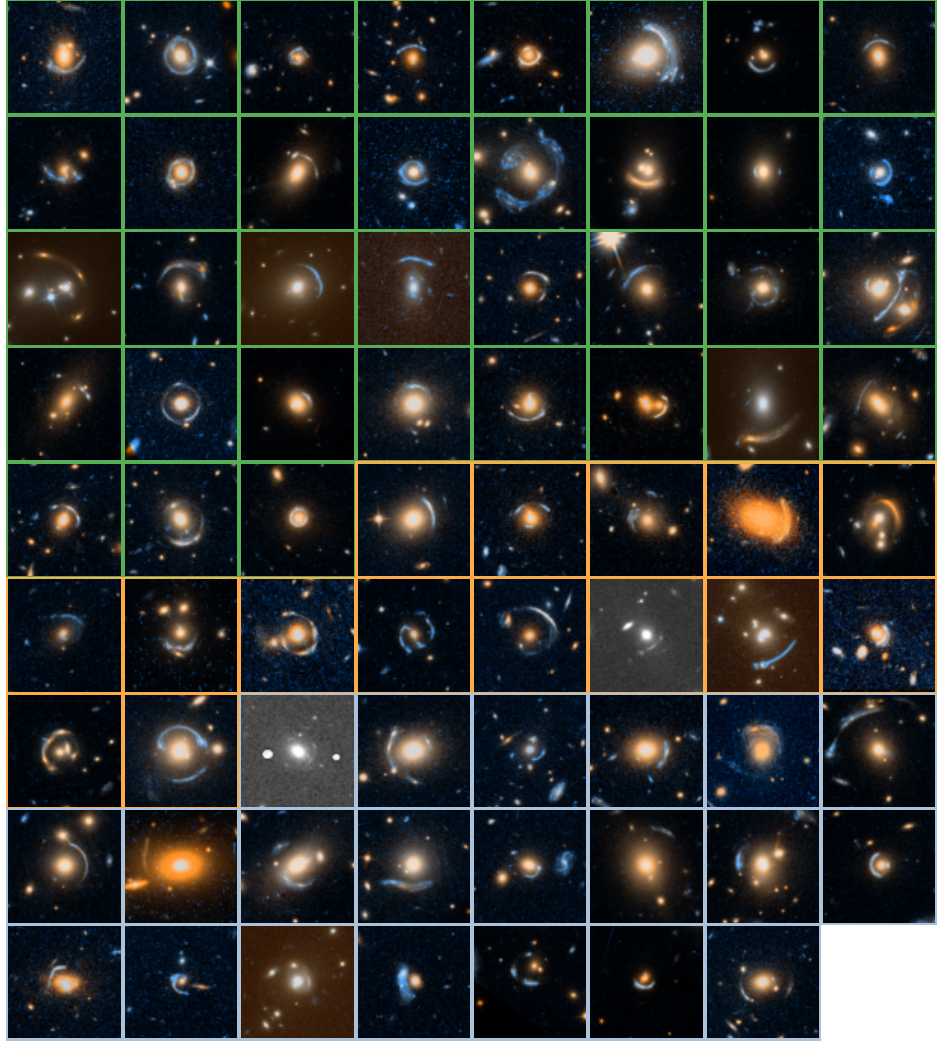}
   \caption{The \HubbleSNAPnumberlenses lenses imaged by HST SNAP program 16773 (PI Glazebrook). Targets bordered in green have confirmed deflector and source redshifts in the AGEL catalogue, targets bordered in orange have either a source or deflector redshift, while a grey border indicates no redshifts in AGEL. All images are $20\arcsec \times 20\arcsec$ with north oriented up and east left. Object names are listed in Table \ref{tbl:hst_snap_megafig}.}
    \label{FIG:HST_SNAP_MEGAFIG}
\end{figure*}

The completed program 16773 obtained WFC3 UVIS/F200LP and IR/F140W imaging and was scheduled \HubbleSNAPnumbertotalvisits visits of the total 150 orbits, (51\%), representing a high visit rate for HST SNAP programs (completion rates for SNAP programs on WFC3 are typically 18\% \footnote{\href{https://www.stsci.edu/files/live/sites/www/files/home/hst/documentation/_documents/UIR_SNAP.pdf}{STScI SNAP User Information Report}}). This high completion rate is attributed to two key factors which maximised the programs schedulability, (i) all sky targets, and (ii) visit durations under 30 minutes. Of those \HubbleSNAPnumbertotalvisits visits, \HubbleSNAPnumbertargets objects were targeted. 2 targets have been spectroscopically confirmed to not be lenses, and 2 other targets are unlikely to be lenses based on their HST imaging. A further 3 targets were affected by guiding issues resulting in partial data loss, however at least one exposure in one band is available. See Table \ref{tbl:hst_snap_comments} for details of these targets. The final tally for the program is therefore \HubbleSNAPnumberlenses lenses, of which \HubbleSnapSomeSpec also have spectroscopy (source or deflector or both). We show all \HubbleSNAPnumberlenses lenses in Figure \ref{FIG:HST_SNAP_MEGAFIG}.

The observing strategy started with $3 \times 300$s in F140W followed by 300s in F200LP. To facilitate better cosmic ray rejection in the F200LP images, and based on the analysis of \cite{Shajib2022} we later updated this strategy to $3 \times 200$s for F140W and $2 \times 300$s for F200LP. 6 targets were observed using the initial strategy, the rest were observed using the updated exposure times (see Table \ref{tbl:hst_snap_comments}). We reduced the images using STScI DrizzlePac to align exposures, correct for background distortion, and remove flagged cosmic rays. To create the combined images, we adjusted the rotation of the stacked image to be in the same orientation across filters. A pixel size of 0.08\arcsec was set to match F200LP to the scaling of F140W.

The F140W filter was chosen to optimise the target S/N, thereby decreasing the exposure time required for lens confirmation. Assuming a $10^{8} M_{\odot}$ source at $z=2$ with a young stellar population lensed with a magnification of $\mu=10$ would result in an H(AB)=26 \citep{Tomczak2014}, which are on the scale of massive star-forming clumps. Depending upon the shape of the spectral energy distribution (SED) the S/N in optical bands will be similar to or less than in the IR channel. For such star-forming sources at $1 < z < 3.5$ with flat SEDs the optical wavelength S/N will be similar, but in instances of reddening due to dust, old stellar population age, or high redshift the IR channel will have a higher S/N. Based on the 3×200s for F140W and 2×300s for F200LP observing strategy used for the majority of the observations, we obtained 5$\sigma$ limiting surface brightnesses of $21.8$ and $26.4$ mag/arcsec in the F140W and F200LP filters respectively.

To cleanly distinguish between source light and foreground/or deflector light it is important to have colour information \citep{Metcalf2019}. In particular, cleanly identifying source counter images is critical for robust lens modelling. Therefore each target was also observed in the wide F200LP filter, which was chosen to maximise S/N and to probe the rest-frame ultraviolet.

\begin{figure*}
    \centering
    \includegraphics[width=\textwidth]{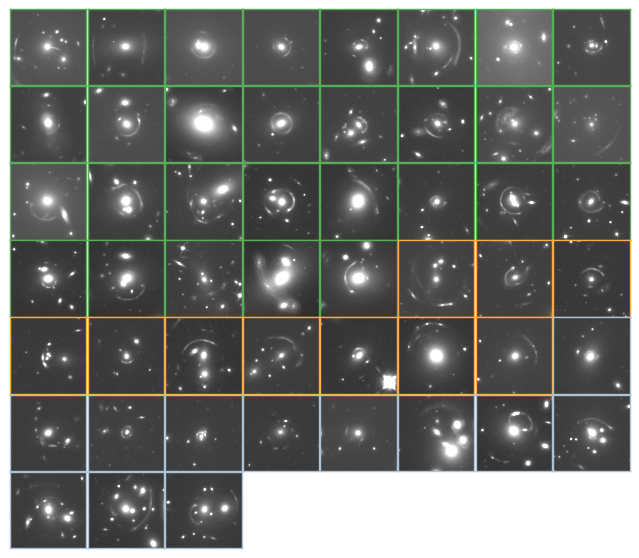}
   \caption{The \HuangGoodImages lenses imaged by HST SNAP program 15867 (PI Huang). Targets bordered in green have confirmed deflector and source redshifts in the AGEL catalogue, targets bordered in orange have either a source or deflector redshift, while a grey border indicates no redshifts in AGEL. All images are $20\arcsec \times 20\arcsec$ with north oriented up and east left. Object names are listed in Table \ref{tbl:huang_snap_fig}. See \cite{Huang2025} for further details of this program.}
    \label{FIG:HUANG_SNAP_FIG}
\end{figure*}

\subsection{ACS GAP Program 17307, PIs Tran and Shajib}
Program 17307 is a (presently ongoing) imaging campaign with ACS in the F606W filter. GAP programs with ACS were introduced by STScI to make use of orbits in which GO and SNAP targets cannot be scheduled. GAP programs therefore require all-sky targets and visit durations of $\lesssim 25$ minutes, and in the past have run for 5 years (compared with SNAP programs which run for 2 years). To increase the scheduability of the observations we use short exposure times of $2 \times 337$s per target. Between the start of observations (November 2023) until the end of 2024 the program has been scheduled \HubbleGAPnumberinclBad orbits. 2 targets suffered data loss so resulted in no usable observations (DESI-271.4849+70.9553 and DES0337-2958), while a third target (DESJ0145+0402) had incorrect sky coordinates so the target was not observed, resulting in \HubbleGAPnumber targets. See Table \ref{tbl:hst_snap_comments} for a summary and Figure \ref{FIG:GAP_HST_FIG} for the images.

The F606W filter was chosen to optimise flux from the typically $z_{\rm source}> 1$ blue, star-forming source galaxies as well as from the rest-frame optical of the $z_{\rm deflector}< 1$ massive quiescent deflectors. The input targets were selected from a series of lens candidate catalogues from the DES survey \citep{Jacobs2019a,Jacobs2019b,J_ODonnell2022}, DELVE survey \citep[DECam Local Volume Exploration;][]{Drlica-Wagner2021, Zaborowski2023}, HSC-SSP \citep[Hyper Suprime-Cam Subaru Strategic Program;][]{Sonnenfeld_2020, Canameras_2021, Shu2022, Chan_2024}, DESI Legacy Imaging Survey \citep{X_Huang2020, X_Huang2021, Storfer2022, Stein2022}, and CASSOWARY survey \citep{Stark2013_CASSOWARY}.

\subsection{WFC3 SNAP Program 15867, PI Huang}
Many AGEL targets have also been observed by SNAP program 15867 \citep[PI Huang;][]{Huang2025} in cycle 27. This program obtained F140W imaging with WFC3/IR for \HuangGoodImages targets\footnote{53 targets were observed in total however 2 observations, DESI-158.2306+75.3149 and DESI-217.0936+03.3000, had failed guide star acquisitions resulting in full data loss. DESI-133.3800+23.3652 also had a failed guide star aquisition but still took useable data.} with 20 minute exposures. See Table \ref{tbl:hst_snap_comments} for a summary, Figure \ref{FIG:HUANG_SNAP_FIG} for the images and \cite{Huang2025} for further details.

\section{Spectroscopic Data}\label{sec:spectroscopic_data}

\begin{figure*}
    \centering
    \includegraphics[width=0.95\textwidth]{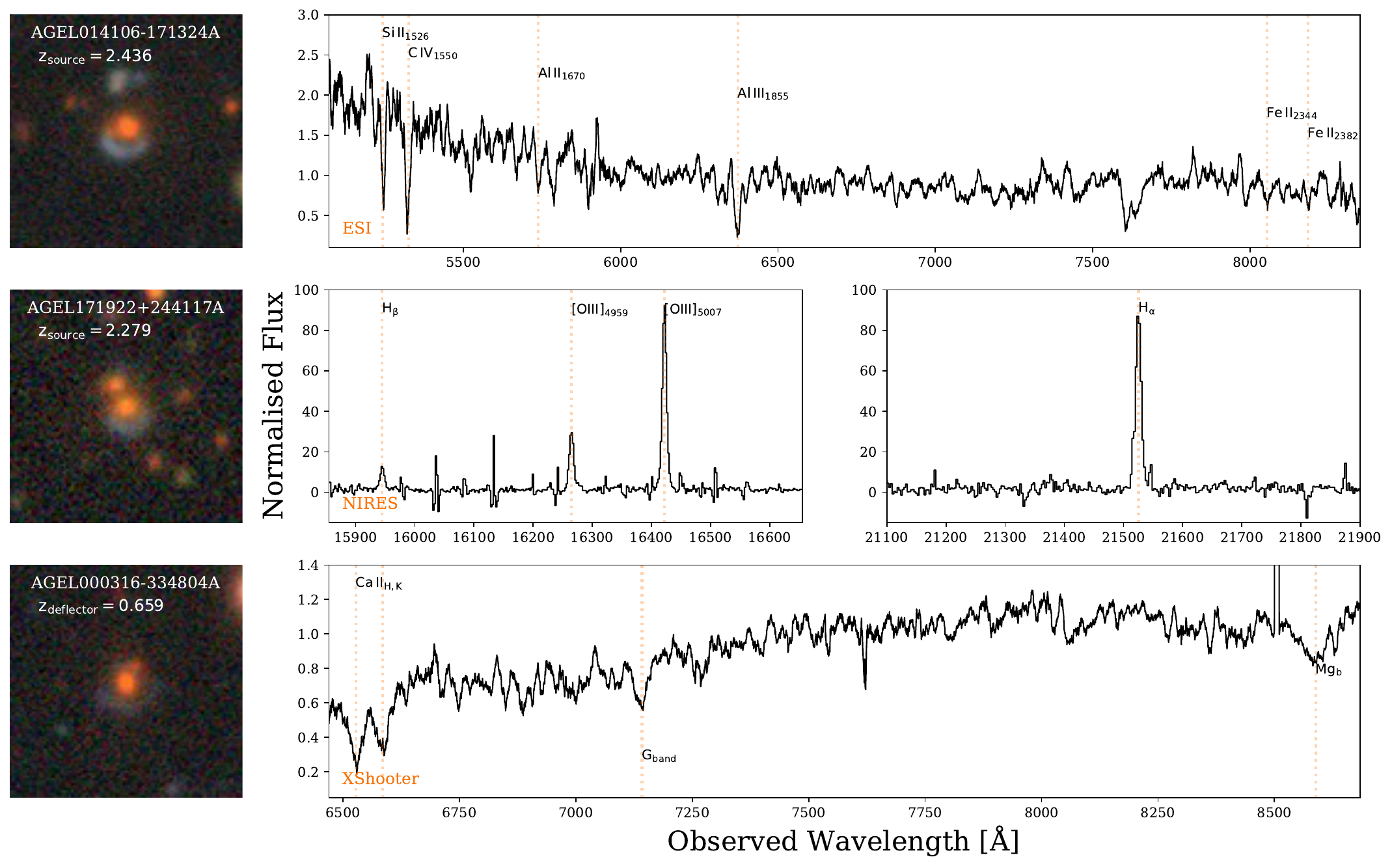}
    \caption{Example AGEL spectra. The top row shows the source spectrum from ESI with the $z=2.436$ redshift measured from ISM absorption features. The middle row  is a $z=2.279$ NIRES source spectrum measured from nebula emission lines. The bottom panel is a $z=0.659$ X-shooter deflector spectrum measured from stellar absorption features. \textit{grz} $25\arcsec \times 25 \arcsec$ DECaLS images are shown in the left column with north oriented up and east left. Key spectral features used in the redshift measurement are shown with dotted orange lines. All spectra have a quality flag of 3.}
    \label{fig:example_spectra_quality3}
\end{figure*}

\begin{figure*}
    \centering
    \includegraphics[width=0.95\textwidth]{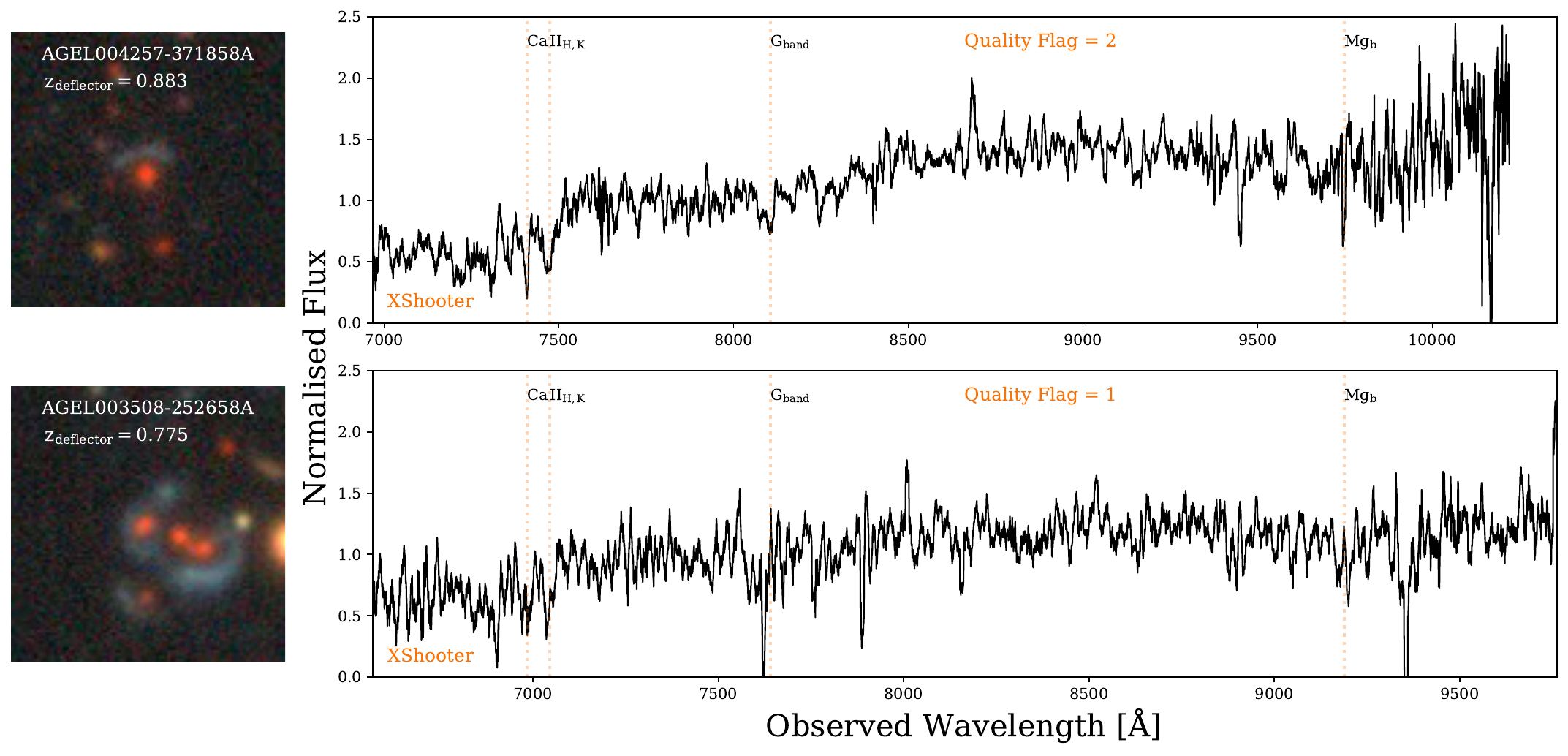}
    \caption{Example AGEL spectra of varying redshift quality flags. The top row is a deflector spectrum from X-shooter with redshift quality flag$=2$, and the bottom row is a deflector spectrum from X-shooter with quality flag$=1$. \textit{grz} $25\arcsec \times 25 \arcsec$ DECaLS images are shown in the left column with north oriented up and east left. Key spectral features used in the redshift measurement are shown with dotted orange lines.}
    \label{fig:example_spectra_quality1-2}
\end{figure*}

AGEL spectroscopic targets were primarily drawn from \cite{Jacobs2019a,Jacobs2019b}, who used convolutional neural networks on the DES and DECaLS fields \citep{Dey2019} to identify lens candidates. Additional targets were drawn from \cite{X_Huang2020, X_Huang2021} who similarly used a neural network approach on the DES and DECaLS fields. We note however that candidates from these catalogues may have been previously identified using other methods, e.g. some candidates had been previously identified by \cite{Diehl2017} using a colour and magnitude selection in the DES catalogues. Therefore, while we include in Table \ref{tbl:redshift_summary} the origin of each system, this may not be the systems's first reported discovery and simply reflects the catalogue we obtained the candidate from.

\subsection{Measuring Redshifts}\label{sec:measuring_redshifts}

To measure the redshifts we used different spectral features depending upon the target's redshift, the observed wavelength coverage and the relative strength of the spectral features. The deflectors are generally massive passive galaxies (or small groups of quiescent galaxies) at $z \sim 0.5$, and therefore redshift confirmation is most easily achieved via the rest-frame optical stellar absorption features  Ca\textsc{ii\textsubscript{h,k}}, Balmer series absorption, Mg\textsubscript{b}, and Na\textsc{d}. We therefore required sufficient continuum S/N over a wide wavelength range to detect these absorption features. To this end, we primarily used Keck/ESI and VLT/X-shooter to obtain deflector redshifts. On the other hand the lensed sources tend to be $z>1$ star-forming galaxies with bright rest-frame optical emission lines\footnote{We confirmed two sources that are $z>1$ quiescent galaxies \citep{Zhuang2022}, with most red sources being dusty star-forming systems.}. This motivated the choice of Keck/NIRES and VLT/Xshooter for source redshifts, which cover rest-frame optical emission lines for $z>1$ sources. This approach also allowed us to measure strong interstellar absorption and stellar wind features such as Si\textsc{ii}, C\textsc{iv}, Fe\textsc{ii}, and Mg\textsc{ii} for many of the sources \citep[e.g.][]{Vasan2023}. See Figures \ref{fig:example_spectra_quality3} and \ref{fig:example_spectra_quality1-2} for example spectra. In addition to the primary spectroscopic observing programs using X-shooter, ESI and NIRES, additional redshifts were measured using Keck/MOSFIRE, Keck/KCWI, and KECK/DEIMOS. We detail the observing strategies and data reduction pipelines used with each instrument in the following subsections.

Tables \ref{tbl:redshift_summary} and \ref{tbl:redshift_summary_extended} summarise all the measured redshifts. Each redshift includes a (subjective) quality flag indicating the reliability of the measurement ($3=$ high, $1=$ low). A redshift flag of 3 indicates a reliable measurement from multiple well-detected features; see Figure \ref{fig:example_spectra_quality3} for example spectra with quality flags $= 3$. A flag of 2 indicates one strong line and a possible weak second line or second line obscured by telluric features, or multiple features in a low S/N spectrum. A flag of 1 was given if only a single feature was observed, or the S/N was low enough to make the redshift measurement uncertain. See Figure \ref{fig:example_spectra_quality1-2} for example spectra with quality flags of 1 and 2. To refine the redshifts we used a mixture of fitting Gaussians to emission and absorption features as well as the online redshifting software MARZ \citep{Hinton2016_MARZ} to fit templates to multiple features.

The redshifts all have formal uncertainties of $\pm 0.0005$, with the following caveats: (a) redshifts with a lower quality flag are less certain (flag 2 will have lower precision, flag 1 lower accuracy and/or precision); (b) redshifts measued from ISM features are likely to be slightly blueshifted from the systemic velocity due to outflows \citep[e.g.][]{Vasan2023}; and (c) depending upon the inclination and lensing geometry the measured redshift may also be convolved with internal galaxy rotation.

\begin{table*}
\resizebox{\textwidth}{!}{
\begin{tabular}{ccl}
\hline
AGEL Name & HST object name & Comment \\
(1) & (2) & (3) \\
\hline
AGEL211515+101153A & DCLS2115+1011 & Ring galaxy \\
AGEL215041+140248A & DECALS 6508900 & Ring galaxy \\
AGEL224400+124540A & DCLS2244+1245 & Red star + blue galaxy \\
AGEL203459+001636A & DECALS 1727734 & Red star + blue galaxy \\
AGEL131330-064211A & DCLS1313-0642 & Red Galaxy + blue blob (not spectroscopically confirmed) \\
AGEL132557+263659A & DCLS1325+2636 & Red Galaxy + blue blob (not spectroscopically confirmed) \\
\hline
\end{tabular}
}\caption{False-positive targets. (1) AGEL object name, (2) object name used in the HST observations, (3) target details.}\label{tbl:false_positive_comments}
\end{table*}

\subsection{Keck/NIRES and ESI}
The primary instruments used for northern and equatorial targets were Keck's ESI \citep[Echellette Spectrograph and Imager;][]{Sheinis2002} and NIRES \citep[Near-Infrared Echellette Spectrometer;][]{Wilson2004}, with ESI used primarily to obtain deflector redshifts and NIRES to measure high ($z>1$) redshift sources.

\NIRESTotalRedshifts redshifts were measured with NIRES using total exposure times of typically 20 minutes (using an ABBA dither pattern). NIRES has a fixed slit of $0.55\arcsec \times 18\arcsec$ providing a mean spectral resolution of $R=2700$ over the wavelength range $0.9-2.45 \micron$ with a pixel scale of 0.15\arcsec/pixel. We reduced the data using the NSX pipeline developed by T.Barlow\footnote{https://sites.astro.caltech.edu/~tb/nsx/} and used the 2D echelle-corrected data produced by the pipeline to extract the 1D spectra.

We obtained \ESITotalRedshifts redshifts using ESI in its echelle mode with the $1.0$\arcsec$ \times 20$\arcsec$ $ slit, providing wavelength coverage of $3900-10900$ {\AA} at a resolving power of $R=4000$. Exposure times varied depending upon the brightness of the target and weather conditions, with typical total exposures between $20 - 80$ minutes. Early observations (programs 2019B\_W226, 2019\_U058) were reduced using the \textsc{ESIRedux} software provided by J. X. Prochaska\footnote{\href{ESIRedux website}{https://www2.keck.hawaii.edu/inst/esi/ESIRedux/}}. Later observations (2020 onwards) were reduced using \textsc{makee} written by T. Barlow\footnote{https://sites.astro.caltech.edu/~tb/makee/}.

\subsection{VLT/X-shooter}
We use the X-shooter instrument \citep{Vernet2011} on the Very Large Telescope to obtain \XhsooterTotalRedshifts redshifts. X-shooter provides slit spectral coverage over the wavelength range $3000-25000$ {\AA}. The data was obtained in programs 0101.A-0577 (PI Glazebrook), 105.20KF.001 (PI Lopez), and 108.22JL.001 (PI Lopez). In program 0101.A-0577 exposure times ranged between 10-40 minutes for deflectors and 40-60 minutes for the sources, with the source and deflector occasionally targeted in the same exposure. For programs 105.20KF.001 and 108.22JL.001, source and deflector spectra were taken separately with typically 30 minute exposures for the deflectors and 60 minute exposures for the sources. We used slit widths of $1.0\arcsec$ (R=5100),  0.9$\arcsec$ (R=8800) and  $0.9\arcsec$ (R=5600), for the UVB ($3000-5600${\AA}), VIS  ($5600-10240${\AA}) and NIR ($10240-24800${\AA}) arms respectively.

The data was reduced using the ESO REFLEX pipeline \citep{Modigliani2010} to obtain 2D flux calibrated spectra in air wavelengths. The flux calibration was performed using standard stars taken during the respective observing night. We then extracted 1D spectra using the \textsc{SpecLib} software package written by C. Schreiber\footnote{https://github.com/cschreib/speclib}.

\subsection{Keck/KCWI}
We used the Keck cosmic Web Imager \citep{Morrissey2018} to obtain \KCWITotalRedshifts spectra. We used the medium slicer with the blue low resolution (BL) grating, which provides a resolution of 2.5 {\AA} FWHM (R$\approx$1800) in the wavelength range 3500–5600 {\AA} in a 16.5\arcsec $\times$ 20.4\arcsec field of view. We reduced the data using the IDL and Python versions of the KCWI data reduction pipeline\footnote{https://kcwi-drp.readthedocs.io/en/latest/} using the standard steps. For complete details of the data reduction process see \cite{Vasan2024}.

\subsection{Keck/MOSFIRE}
\MOSFIRETotalRedshifts targets were observed using Keck's MOSFIRE instrument \citep[Multi-Object Spectrometer For Infra-Red Exploration;][]{Mclean2012} as filler targets for program 2022A\_W223 (PI Nanayakkara). The observations on AGEL targets used the $1.0
\arcsec \times 60\arcsec$ longslit in the H-band (R=3600 over $1.47-1.80 \micron$) with additional exposures in the K-band (R=3610 over $2.03-2.40 \micron$) if needed. The data was reduced using the publicly available pipeline\footnote{http://keck-datareductionpipelines.github.io/MosfireDRP/} using the procedure outlined in \citet{Nanayakkara2016}. For more information about the observing run, see \citet{Antwi-Danso2023}.

\subsection{Keck/DEIMOS}
We targeted \DEIMOSTotalRedshifts system, AGEL235934+020824A, with the Keck/DEIMOS instrument (2022B\_U099, PI Jeltema). We used two different instrument configurations. The first configuration targeted the central galaxy as well as candidate cluster galaxies in the field with the 1200G grating, yielding a spectral range of roughly $5200-8000${\AA}. The second configuration targeted the lensed source as well as candidate cluster galaxies, using the 600ZD grating for maximum wavelength coverage. All data from this night were reduced with PypeIt version 1.11.0 \citep{pypeit:zenodo,pypeit:joss_pub}. While the spectrum of the deflector galaxy yielded a highly confident redshift, the source spectrum detected no features. The lensed source was later targeted by NIRES (2023B\_U039, PI Jeltema), obtaining a confident redshift.

\begin{figure}
    \centering
    \includegraphics[width=0.95\columnwidth]{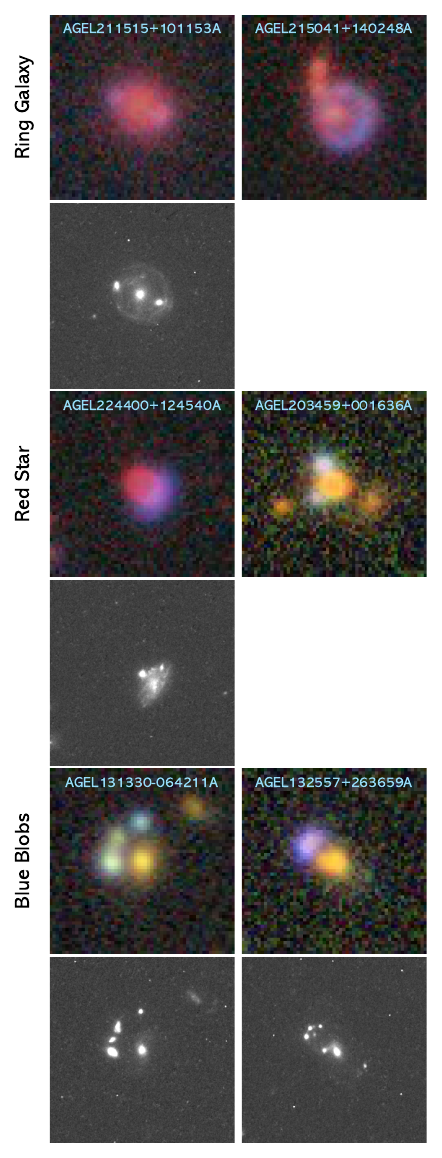}
    \caption{The 6 candidates found to not be lenses, showing the DECaLS \textit{grz} on the top and the F200LP HST image (if available) beneath it. The \NotLensNum false-positives fall into 3 categories: ring galaxies, red stars, and blue blobs. All images are $15 \times 15 \arcsec$ and are oriented with north up and east left.}
    \label{fig:not_lenses_fig}
\end{figure}

\begin{figure}
    \centering
    \includegraphics[width=\columnwidth]{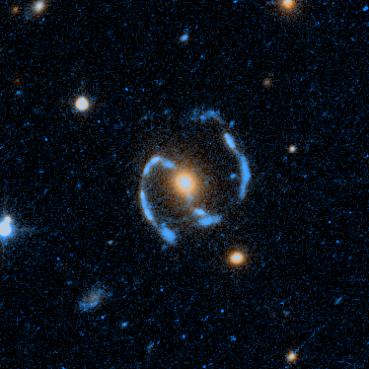}
    \caption{AGEL085413-042409A has a morphology reminiscent of a barred spiral, but is comprised of two $z=2.393$ galaxies multiply lensed by a $z=0.856$ galaxy deflector. The image is $20\arcsec \times 20\arcsec$ with north oriented up and east left.}
    \label{fig:masquerading_spiral}
\end{figure}

\section{Sample Characteristics}\label{sec:sample_characteristics}

\subsection{Impostors}\label{sec:imposters}

The AGEL lens confirmation campaign has achieved a high success rate (\SuccessRate\%), however during the course of the survey we have identified \NotLensNum lens candidates which, upon spectroscopic and/or photometric followup, were identified as false-positives. Identifying the common causes of impostors is important to ensuring high fidelity samples in upcoming large area surveys such as LSST and Euclid. We therefore present these false positives in Figure \ref{fig:not_lenses_fig}, and briefly discuss their morphologies.

The \NotLensNum false-positives fall into 3 categories: (a) ring galaxies, (b) a red star in front of blue galaxies, and (c) red galaxies next to irregular blue blobs (see Table \ref{tbl:false_positive_comments}). The first category, ring galaxies, need spectroscopic redshifts to identify as shown by the still-convincing HST imaging for AGEL211515+101153 in Figure \ref{fig:not_lenses_fig}.  The second catagory, red stars, could be reduced by cross-matching candidate lists with large catalogues of stars. For example, SDSS uses SED fitting to multi-band photometry to classify targets as stars, galaxies or quasars. However this approach would not completely eliminate red stars as in the example of AGEL224400+124540A which is classified as a galaxy in SDSS, likely due to the light from the blue galaxy behind it contaminating the photometry. Finally, red galaxies next to irregular blue blobs are difficult to eliminate without higher resolution imaging. However upon close inspection the 3 blobs of AGEL131330-064211A have slightly different colours in the DECaLS \textit{grz} imaging leaving it unlikely it is the same object multiply lensed. To rule out false positives, next generation neural network lens finders should aim to take advantage of multiband photometry to (i) accurately identify galaxy deflectors from red stars, and (ii) find candidate arcs with identical colours. High resolution multi-band images from Euclid should therefore significantly improve future high-fidelity lens searches.

While high resolution imaging can usually confirm the lensing nature of candidates, it is not always sufficient and occasionally spectra is also required. AGEL085413-042409A (Figure \ref{fig:masquerading_spiral}) has a morphology reminiscent of spiral arms, and as a result HST imaging alone did not definitively confirm whether AGEL085413-042409A is a lens or whether the blue arcs are barred spiral arms or features of a merger. Integral field spectroscopy with Keck/KCWI confirmed the blue ``arms'' are two $z=2.393$ galaxies, both multiply lensed by a $z=0.856$ galaxy deflector. %

\subsection{Comparison to other surveys}\label{sec:comparison_other_surveys}

We explore how the AGEL sample compares in redshift distribution from previous lens samples. Following \citealt{Tran2022}, in the following comparisons we include all targets which satisfy the following criterion: spectroscopic redshifts for the foreground deflector and/or background source which have a quality flag $\geq 2$ (see Section \ref{sec:measuring_redshifts} and Figures \ref{fig:example_spectra_quality3} and \ref{fig:example_spectra_quality1-2} for definitions of the quality flag). This criterion removes \TotalRedshiftsLowQuality targets.

\begin{figure}
    \centering
    \includegraphics[width=\columnwidth]{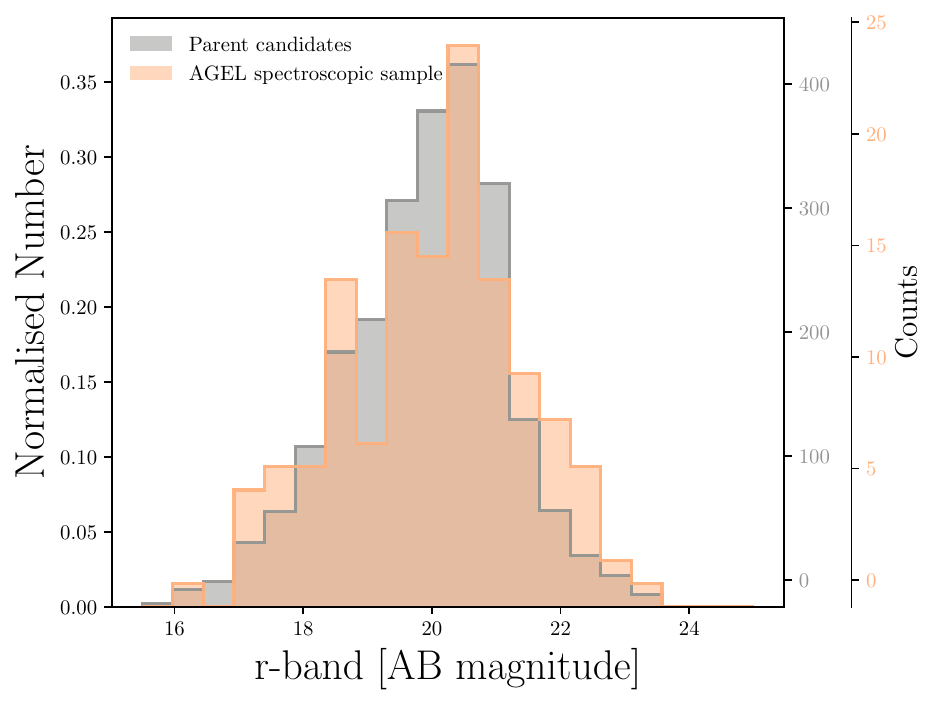}
    \caption{Number density histogram of deflector r-band magnitudes. The AGEL parent candidate sample from \cite{Jacobs2019a,Jacobs2019b,X_Huang2020,X_Huang2021} is shown in grey, with the spectroscopic sample in orange. Based on 1000 realisations of the parent sample, we find the two distributions statistically consistent.}
    \label{fig:rband_parent_spec_sample}
\end{figure}

As discussed previously in \citealt{Tran2022}, candidate lenses were prioritised for spectroscopic observations if (a) they were at suitable RA and DEC for the allocated observations, (b) they had high resolution imaging, and (c) they appeared to be lenses in visual inspection of available imaging. It is therefore difficult to directly compare the resulting AGEL sample to surveys such as SLACS, BELLS, and CSWA and SL2S with comparatively well defined selection functions. As shown by Figure \ref{fig:rband_parent_spec_sample}, the deflector r-band magnitudes of the spectroscopic sample differ slightly to the parent candidate samples of \cite{Jacobs2019a,Jacobs2019b} and \cite{X_Huang2020,X_Huang2021}. Comparing the DR2 sample to 1000 random realisations of the parent sample, we find the DR2 sample is slightly broader than parent distribution (although the standard deviations are consistent within 2$\sigma$) and with means consistent within $1\sigma$. This excess in the spectroscopic sample to fainter deflector magnitudes likely reflects the $z > 0.5$ science cases in mind for AGEL deflectors \citep[e.g.][]{Sahu2024,Barone2024}. However, for the purposes of the following comparison, we consider the AGEL sample as representative of the total candidates found by neural networks. In the following section we consider how differences in the AGEL sample to literature samples may be reflective of differences in identification methods.

\begin{figure*}
    \centering
    \includegraphics[width=\textwidth]{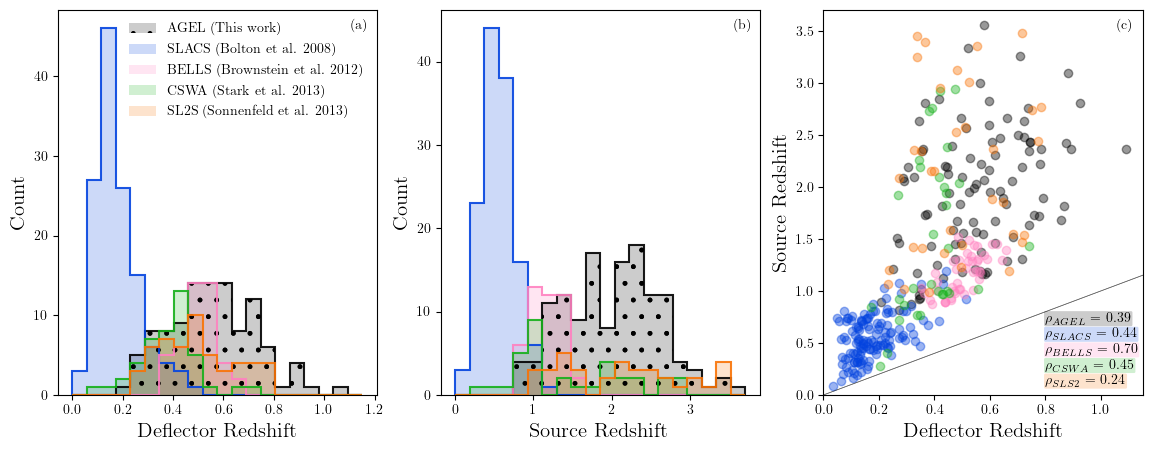}
    \caption{Comparison of the AGEL spectroscopic sample to other lensing samples. Panel (a) shows the deflector redshifts, (b) the source redshifts (all redshifts spectroscopically confirmed). The AGEL sample spans to higher source and deflector redshifts than most previous samples. Panel (c) shows the relationship between source and deflector redshifts, with the Spearman rank-order correlation coefficient for each sample annotated in the bottom right corner. All the samples have p-value$<<0.05$ except for SL2S, implying that SL2S is the only sample in which the source redshift is independent to the deflector redshift. The grey line in panel (c) shows the one-to-one relation.}
    \label{fig:redshift_histogram}
\end{figure*}

Figure \ref{fig:redshift_histogram} shows the deflector and source redshift distributions for the AGEL sample (candidates identified in DES and DECaLS imaging via neural networks), SLACS and BELLS (identified in SDSS single-fibre spectroscopy), CASSOWARY (identified via blue sources close to luminous red galaxies in SDSS imaging), and SL2S (blue sources close to luminous red galaxies in CFHT imaging). The different selection methods evidently lead to different deflector and source distributions.

Firstly, the redshift distributions of the fibre based search surveys are intrinsically linked to the aperture of the fibre (3$\arcsec$ diameter for SLACS and 2$\arcsec$ diameter for BELLS). The Einstein radius $\theta_E$ of a single isothermal sphere is:

\begin{equation}
    \theta_E = 4 \pi \frac{\sigma_v^2}{c^2} \frac{D_{ds}}{D_s}
\end{equation}

For $D_{ds}$ the angular diameter distance between deflector and source, $D_s$ the distance between observer and source, and $\sigma_v$ the velocity dispersion of the deflector. Therefore, for fixed deflector velocity dispersion, $\theta_E \propto \frac{D_{ds}}{D_s}$. We can see this in Figure \ref{fig:redshift_histogram}c where the source redshifts tend to be more correlated to the deflector redshifts in the SLACS and BELLS samples (Spearman coefficients of $\rho = 0.44 \ \text{and}\ 0.70$ respectively) than the image-based search samples (c.f. $\rho=0.38$, $0.45$ and $0.24$ for AGEL, CASSOWARY and SL2S).

By virtue of the deeper sensitivities of DES and DECaLS imaging ($r_\mathrm{DES,DECaLS} = 23.4$ compared to $r_\mathrm{SDSS} = 22.70$) our sample spans higher redshifts for both deflectors and sources than the SDSS based surveys (SLACS, BELLS, CSWA).

The SL2S and AGEL deflector distributions are most similar. CFHTS has slightly deeper imaging than DECaLS ($r_\mathrm{CFHTLS}=24.5$), however DECaLS covers a significantly larger area ($\sim 9000$ deg$^2$ compared to CFHTS $\sim 170$ deg$^2$). Noticably, although AGEL has more $z_{\rm deflector} > 0.8$ (possibly due to the larger search area), the SL2S source redshifts skew higher, likely due to the deeper imaging. Similarly to AGEL, SL2S used a range of telescopes and instruments for spectroscopic follow-up. Therefore whether a redshift was securely measured (both in AGEL and SL2S) depends upon the wavelength coverage and sensitivity of the various instruments.

\begin{figure}
    \centering
    \includegraphics[width=\linewidth]{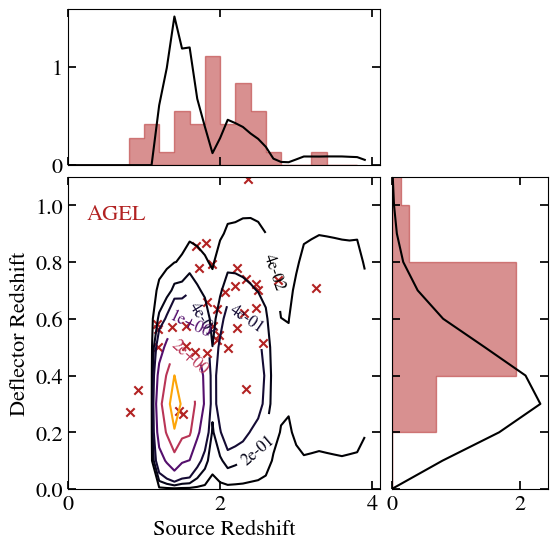}
    \caption{Comparison between the redshift distribution of the AGEL galaxy-scale subsample (red crosses and histograms) and predictions from the \textsc{galess} analytical model \citep[contours and black solid line;][]{Ferrami2024} . AGEL systems skew to higher redshift deflectors than expected by the model.}
    \label{fig:expected_distribution}
\end{figure}

\begin{table*}
\resizebox{\textwidth}{!}{
\begin{tabular}{clcccc}
\hline
AGEL Name & HST object name & HST program ID & $z_{A}$ & $z_{B}$ & $z_{DE}$ \\
(1) & (2) & (3) & (4) & (5) & (6)\\ \hline
AGEL150745+052256 & DCLS1507+0522 & 16773 & 2.163 & 2.591 & 0.594 \\
AGEL080820+103142 & DESI-122.0852+10.5284 & 15867 & 1.303 & 1.452 & 0.475 \\
AGEL035346-170639 & DCLS0353-1706 & 16773 & 1.674 & 1.46 & 0.617 \\
AGEL013442+043350 & DESI-023.6765+04.5639 & 15867 & 1.568 & 2.03 & 0.55 \\
AGEL144149+144121 & DESI-220.4549+14.6891 & 15867 & 1.433 & 2.34 & 0.741 \\
AGEL024303-000600 & DESJ0243-0006 & 16773 & 1.729 & 0.506 & 0.367 \\
\hline
\end{tabular}
}\caption{Galaxy-scale double source lenses. The columns are as follows: (1) The object label used in the AGEL catalogue. (2) Object name  used in the HST observations. (3) ID of the observing program the HST images came from; ID:16773 (PI Glazebrook) has filters F140W and F200LP and ID:15867 (PI Huang) has just the F140W filter. (4) and (5) show the redshifts of the arcs as labelled in Figure \ref{fig:multi_source_lenses}. (6) Deflector redshift.}\label{tbl:multi_source_lenses}
\end{table*}

\begin{figure*}
    \includegraphics[width=\textwidth]{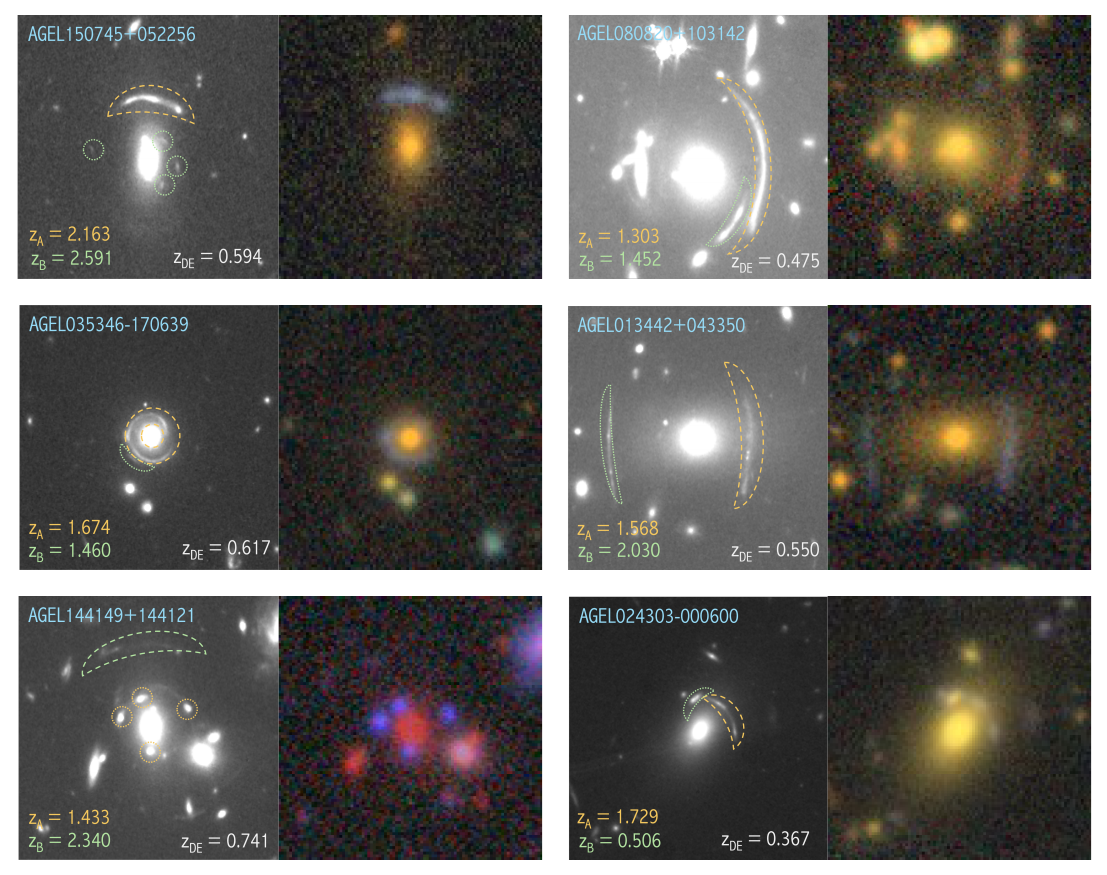}
    \caption{Galaxy-scale multi-source lenses, with the HST/F140W image in the left column and DECaLS \textit{grz}  in the right column. Arc A is highlighted by the orange dashed shapes and arc B by the green dotted shapes. The redshifts are labelled in the HST panels. All images are $25 \times 25 \arcsec$ and are oriented with north up and east left.}
    \label{fig:multi_source_lenses}
\end{figure*}

The number density of galaxy-scale lenses detected in wide-area surveys is sensitive to cosmology. From the AGEL redshift sample we select a subset of \NumGalaxyScaleLenses that have (i) both source and deflector redshifts, and (ii) have a single-galaxy deflector based on the DECaLS grz imaging. We compare the redshift distribution of this galaxy-scale subset to the distribution predicted by the analytical model \textsc{galess}\footnote{\url{https://github.com/Ferr013/GALESS/}} \citep{Ferrami2024}, assuming the limiting magnitudes of the DECaLS imaging ($g=24.0$, $r=23.4$ and $z=22.5$). For the model we limit the source redshifts to the intervals in which at least two emission lines out of $H_\alpha$, $H_\beta$, $[O_{III}]$, and $[O_{II}]$ can be detected in NIR infrared spectroscopy. The resulting number density distribution is shown in Figure \ref{fig:expected_distribution}. From Figure \ref{fig:expected_distribution}, AGEL deflectors skew to higher redshifts ($z \sim 0.6$) than predicted by the model (which peaks at $z\sim 0.3$). This may be related to the subtle skew in AGEL deflectors to fainter r-band magnitudes than the parent catalogue (Figure \ref{fig:rband_parent_spec_sample}), potentially reflecting the team's science preference towards identifying higher redshift deflectors.

\subsection{Double Source Plane Lenses}\label{sec:double_source_lenses}
A key science case driving the need for large lens samples with high-resolution imaging is cosmology via double source plane galaxy-scale lenses \citep[gravitational lens cosmography;][]{Blandford_Narayan1992, Treu2010_annual_review}.

\begin{figure}
    \includegraphics[width=\linewidth]{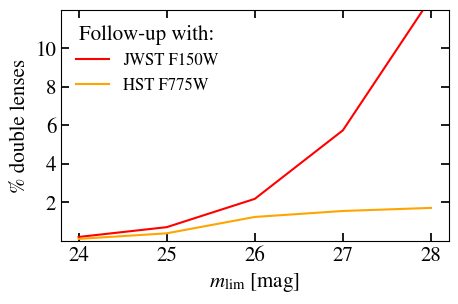}
    \caption{Prediction on the fraction of galaxy-scale lenses originally identified in DECaLS photometry that are expected to show additional lensed arcs if observed in F150W (red line) and F775W (orange line) filter observations, based upon the limiting magnitude ($m_{\rm lim}$) of those observations. Prediction based on semi-analytical model \textsc{galess} \citep{Ferrami2024}}.
    \label{fig:frac_double_lenses}
\end{figure}

The ideal system for gravitational lens cosmography is one that has (i) a deflector with a simple mass distribution, and (ii) numerous, multiply lensed arcs at a large range of redshifts. Galaxy-scale lenses satisfy the first criterion, with mass profiles that are close to single isothermal spheres. However, galaxy-scale systems rarely lens more than one source \cite[the current record is 3;][]{Collett_Smith2020}, so seldom satisfy the second criterion. In theory, gravitational lens cosmography is possible from single source systems \cite[e.g.][]{Biesiada2010}, but the cosmological parameters are highly degenerate with the choice of the deflector mass density profile. A second multiply lensed source provides additional constraints on the mass distribution and therefore reduces this degeneracy \citep{Collett_Auger2014,Smith_Collett2021}. Furthermore the constraining power of each system depends upon the relative redshifts of the sources, with a low separation lens--source combined with a second high redshift source proving ideal \citep{Collett2012}. Therefore, searching for and building a sample of multi-source galaxy-scale is valuable, as even just $\sim10$ systems would provide cosmological constraints with a precision comparable (and orthogonal) to all current supernovae and galaxy surveys \citep{Collett2012}.

We present \NumDoubleSourceGalaxyScaleInBothHSTsnapPrograms double source plane lenses  in the AGEL survey (Table \ref{tbl:multi_source_lenses} and Figure \ref{fig:multi_source_lenses}). These targets were identified by visual inspection of HST imaging and redshifts confirmed spectroscopically. The importance of high resolution imaging to (a) identify faint secondary arcs and (b) identify any possible satellite or neighbouring galaxies is illustrated in Figure \ref{fig:multi_source_lenses}.

The probability of a galaxy acting as a gravitational lens scales with the velocity dispersion ($\sigma$) of the galaxy: $P(\rm{lens}) \sim \theta_{\rm{E}}^2 \sim \sigma^4$ where $\theta_{\rm{E}}$ is the lensing cross section. Hence a sample of galaxy-scale lenses is biased to be more massive than randomly drawn galaxies. Consequently, the conditional probability of observing a second lensed image behind a galaxy-scale lens is higher than the probability of finding a single lens in the first place. Furthermore, AGEL targets were selected from ground-based surveys so the seeing limit in the images leads to only identifying lenses with large Einstein radii. Therefore, just the higher spatial resolution of space telescope images increases the chances of identifying additional arcs with smaller Einstein radii. Combined with the typically deeper sensitivities of space telescope images results in a significantly higher probability of identifying secondary arcs given a first.

To quantify the conditional probability of finding additional arcs, we use the analytic model described in \cite{Gavazzi2008} and implemented in \textsc{galess} \citep{Ferrami2024} to predict the fraction of AGEL galaxy-scale lenses we expect to have additional arcs detectable in HST WFC3/F775W and JWST NIRCam/F150W imaging (Figure \ref{fig:frac_double_lenses}). At limiting magnitudes fainter than 26 we see a significant increase in the fraction of double-source plane lenses detected in the redder F150W filter, as the redder filter is sensitive to higher redshift arcs. Future JWST observations with redder bands therefore have the potential to significantly increase the sample of known double-source plane lenses. In HST programs 16773 and 15867 (with F140W AB limiting magnitudes of approximately 26), we count a total \NumDoubleSourceGalaxyScaleInBothHSTsnapPrograms double-source galaxy-scale lenses out of \NumGalaxyScaleInBothHSTsnapPrograms (5 \%), slightly higher but similar to the 2 \% predicted by the model for the similar F150W filter. However, the local environment of these lenses is not well constrained- these targets were identified as ``galaxy-scale'' by visual inspection of the available imaging. While the arcs appear deflected by a single galaxy, that single galaxy may itself be part of a larger group.

\section{Summary}

The AGEL survey is an ongoing effort to create a large sample of spectroscopically confirmed gravitational lenses with high-resolution imaging. Thus far we have obtained high-resolution HST imaging of \TotalGoodHSTimagesLenses targets, as well as \TotalNumMeasuredRedshifts redshifts associated with \TotalLensSystems lens systems. Compared to previous samples, AGEL lenses span higher source and deflector redshifts. %

Due to the flexible neural network search algorithm, there is a large diversity of targets within AGEL. Deflectors range from individual galaxies all the way up to clusters, and sources span from highly star-forming to long quiescent. As a result, AGEL targets benefit a large range of science topics, including the baryon cycle \citep{Vasan2023,Barone2024}, the build-up of chemical elements \citep{Zhuang2022}, and mass assembly \citep{Sahu2024,Sheu2024}. The insights learnt from the AGEL sample will help to refine lens searches on future large-area surveys.

\begin{table*}
\resizebox{\textwidth}{!}{
\begin{tabular}{c|ccc|ccc|cccc}
\hline
Object Name & $z_{\rm{source}}$ & Source flag & Source Proposal ID & $z_{\rm{deflector}}$ & Deflector flag & Deflector Proposal ID & HST Imaging & Candidate Catalogue & RA & Dec \\
(1) & (2) & (3) & (4) & (5) & (6) & (7) & (8) & (9) & (10) & (11)\\\hline\hline
AGEL000224-350716A & 1.549 & 3.0 & LOPEZ\_105.20KF.001\_XShooter & 0.576 & 3.0 & LOPEZ\_105.20KF.001\_XShooter &  & Jacobs et al. 2019a,b & 0.59845 & -35.12122 \\
AGEL000316-334804A & 1.834 & 3.0 & LOPEZ\_105.20KF.001\_XShooter & 0.659 & 3.0 & LOPEZ\_105.20KF.001\_XShooter & 16773, Glazebrook & Jacobs et al. 2019a,b & 0.81825 & -33.80120 \\
AGEL000645-442950A & 2.099 & 3.0 & LOPEZ\_105.20KF.001\_XShooter & 0.498 & 3.0 & LOPEZ\_105.20KF.001\_XShooter & 16773, Glazebrook & Jacobs et al. 2019a,b & 1.68592 & -44.49735 \\
AGEL000729-443446A & 1.009 & 3.0 & LOPEZ\_108.22JL.001\_XShooter &  &  &  & 16773, Glazebrook & Jacobs et al. 2019a,b & 1.87201 & -44.57949 \\
AGEL001310+004004A & 2.071 & 1.0 & JONES\_2019B\_U058\_ESI & 0.693 & 3.0 & JONES\_2019B\_U058\_ESI & 16773, Glazebrook & Jacobs et al. 2019a,b & 3.29016 & 0.66767 \\
AGEL001424+004145A & 1.374 & 3.0 & GLAZEBROOK\_0101.A-0577\_XShooter & 0.569 & 3.0 & GLAZEBROOK\_0101.A-0577\_XShooter &  & Jacobs et al. 2019a,b & 3.60116 & 0.69596 \\
AGEL001702-100911A & 1.831 & 2.0 & KACPRZAK\_2022B\_W211\_NIRES &  &  &  & 15867, Huang & Huang et al. 2021 & 4.25640 & -10.15300 \\
AGEL002527+101107A & 2.397 & 3.0 & KACPRZAK\_2020B\_W127\_NIRES & 0.463\^ &  & SDSS & 15867, Huang & Huang et al. 2021 & 6.36430 & 10.18530 \\
AGEL002700-041324A & 1.465 & 2.0 & KACPRZAK\_2019B\_W226\_ESI & 0.495\^ &  & SDSS &  & Jacobs et al. 2019a,b & 6.75025 & -4.22321 \\
AGEL003508-252658A &  &  &  & 0.775 & 1.0 & LOPEZ\_105.20KF.001\_XShooter &  & Jacobs et al. 2019a,b & 8.78153 & -25.44932 \\
AGEL003727-413150A & 3.257 & 2.0 & GLAZEBROOK\_0101.A-0577\_XShooter & 0.708 & 3.0 & GLAZEBROOK\_0101.A-0577\_XShooter &  & Jacobs et al. 2019a,b & 9.36280 & -41.53054 \\
AGEL004144-233905A & 2.203 & 3.0 & JELTEMA\_2024B\_U225\_KCWI & 0.439 & 3.0 & JELTEMA\_2024B\_U225\_KCWI &  & Huang et al. 2021 & 10.43410 & -23.65150 \\
AGEL004257-371858A & 3.094 & 3.0 & LOPEZ\_105.20KF.001\_XShooter & 0.883 & 2.0 & LOPEZ\_105.20KF.001\_XShooter & 16773, Glazebrook & Jacobs et al. 2019a,b & 10.73881 & -37.31623 \\
AGEL004827+031117A & 2.367 & 2.0 & JONES\_2019B\_U058\_ESI & 0.357\^ &  & SDSS &  & Jacobs et al. 2019a,b & 12.11340 & 3.18808 \\
AGEL010128-334319A & 1.167 & 3.0 & GLAZEBROOK\_0101.A-0577\_XShooter & 0.581 & 2.0 & GLAZEBROOK\_0101.A-0577\_XShooter &  & Jacobs et al. 2019a,b & 15.36604 & -33.72201 \\
AGEL010158-491738A & 2.632 & 2.0 & LOPEZ\_108.22JL.001\_XShooter &  &  &  & 16773, Glazebrook & Jacobs et al. 2019a,b & 15.49182 & -49.29394 \\
AGEL010238+015857A & 1.816 & 2.0 & JONES\_2019B\_U058\_ESI & 0.867 & 2.0 & JONES\_2019B\_U058\_ESI & 16773, Glazebrook & Jacobs et al. 2019a,b & 15.65959 & 1.98243 \\
AGEL010257-291122A & 0.815 & 3.0 & LOPEZ\_108.22JL.001\_XShooter & 0.273\^ &  & Blake et al. 2016 &  & Jacobs et al. 2019a,b & 15.73954 & -29.18939 \\
AGEL011759-052718A & 2.066 & 3.0 & KACPRZAK\_2020B\_W127\_NIRES & 0.579 & 3.0 & JONES\_2019B\_U058\_ESI & 17307, Tran & Jacobs et al. 2019a,b & 19.49477 & -5.45492 \\
AGEL012429-291856A & 1.606 & 3.0 & LOPEZ\_108.22JL.001\_XShooter &  &  &  & 16773, Glazebrook & Jacobs et al. 2019a,b & 21.11886 & -29.31561 \\
AGEL012453-144303A & 1.825 & 1.0 & KACPRZAK\_2020B\_W127\_NIRES & 0.478 & 3.0 & JONES\_2019B\_U058\_ESI &  & Jacobs et al. 2019a,b & 21.22109 & -14.71738 \\
AGEL013003-374458A & 2.282 & 3.0 & LOPEZ\_105.20KF.001\_XShooter &  &  &  & 17307, Tran & Jacobs et al. 2019a,b & 22.51201 & -37.74938 \\
AGEL013355-643413A & 2.093 & 3.0 & LOPEZ\_105.20KF.001\_XShooter & 0.326 & 3.0 & LOPEZ\_105.20KF.001\_XShooter & 16773, Glazebrook & Jacobs et al. 2019a,b & 23.47773 & -64.57028 \\
AGEL013442+043350A & 1.568 & 3.0 & KACPRZAK\_2020B\_W127\_NIRES & 0.551\^ &  & SDSS & 15867, Huang & Huang et al. 2020 & 23.67650 & 4.56390 \\
AGEL013442+043350B & 2.031 & 3.0 & KACPRZAK\_2020B\_W127\_NIRES & 0.551\^ &  & SDSS & 15867, Huang & Huang et al. 2020 & 23.67650 & 4.56390 \\
AGEL013639+000818A & 2.629 & 3.0 & JONES\_2020B\_U044\_NIRES & 0.344 & 3.0 & JONES\_2020B\_U044\_ESI & 15867, Huang & Jacobs et al. 2019a,b & 24.16310 & 0.13836 \\
AGEL013719-083056A & 2.997 & 3.0 & JONES\_2019B\_U058\_ESI & 0.563 & 3.0 & JONES\_2019B\_U058\_ESI &  & Jacobs et al. 2019a,b & 24.32850 & -8.51552 \\
AGEL014106-171324A & 2.436 & 3.0 & JONES\_2019B\_U058\_ESI & 0.609 & 3.0 & JONES\_2019B\_U058\_ESI & 16773, Glazebrook & Jacobs et al. 2019a,b & 25.27558 & -17.22326 \\
AGEL014235-164818A & 2.309 & 3.0 & KACPRZAK\_2020B\_W127\_NIRES & 0.619 & 2.0 & JONES\_2020B\_U044\_ESI &  & Jacobs et al. 2019a,b & 25.64570 & -16.80487 \\
AGEL014253-183116A & 2.470 & 3.0 & GLAZEBROOK\_0101.A-0577\_XShooter & 0.637 & 3.0 & GLAZEBROOK\_0101.A-0577\_XShooter & 16773, Glazebrook & Jacobs et al. 2019a,b & 25.72030 & -18.52105 \\
AGEL014327-085021A & 2.755 & 3.0 & KACPRZAK\_2020B\_W127\_NIRES & 0.737 & 3.0 & KACPRZAK\_2019B\_W226\_ESI & 17307, Tran & Jacobs et al. 2019a,b & 25.86222 & -8.83925 \\
AGEL014504-045551A & 1.960 & 3.0 & JONES\_2020B\_U044\_NIRES & 0.635 & 3.0 & JONES\_2020B\_U044\_ESI & 17307, Tran & Jacobs et al. 2019a,b & 26.26791 & -4.93084 \\
AGEL014556+040229A & 2.361 & 3.0 & JONES\_2020B\_U044\_NIRES & 0.784 & 3.0 & JONES\_2020B\_U044\_ESI & 17307, Tran & Jacobs et al. 2019a,b & 26.48478 & 4.04139 \\
AGEL015009-030438A & 1.390 & 1.0 & KACPRZAK\_2020B\_W127\_NIRES &  &  &  &  & Jacobs et al. 2019a,b & 27.53794 & -3.07730 \\
AGEL015153-144825A & 2.258 & 3.0 & KACPRZAK\_2021B\_W242\_NIRES &  &  &  &  & Huang et al. 2020 & 27.97230 & -14.80690 \\
AGEL015643-101100A & 2.486 & 3.0 & KACPRZAK\_2021B\_W242\_NIRES &  &  &  &  & Jacobs et al. 2019a,b & 29.17803 & -10.18339 \\
AGEL020613-011417A & 1.303 & 3.0 & JONES\_2020B\_U044\_NIRES & 0.714\^ &  & SDSS & 16773, Glazebrook & Jacobs et al. 2019a,b & 31.55611 & -1.23817 \\
AGEL020707-272645A & 1.679 & 3.0 & LOPEZ\_105.20KF.001\_XShooter &  &  &  &  & Jacobs et al. 2019a,b & 31.77773 & -27.44577 \\
AGEL020707-272645B & 1.677 & 3.0 & LOPEZ\_105.20KF.001\_XShooter &  &  &  &  & Jacobs et al. 2019a,b & 31.77773 & -27.44577 \\
AGEL021225-085211A & 2.202 & 1.0 & KACPRZAK\_2020B\_W127\_NIRES & 0.759\^ &  & SDSS & 16773, Glazebrook & Jacobs et al. 2019a,b & 33.10507 & -8.86967 \\
AGEL022709-471856A &  &  &  & 0.603 & 3.0 & GLAZEBROOK\_0101.A-0577\_XShooter & 16773, Glazebrook & Jacobs et al. 2019a,b & 36.78734 & -47.31550 \\
AGEL022931-290816A & 1.679 & 3.0 & LOPEZ\_105.20KF.001\_XShooter & 0.857 & 2.0 & LOPEZ\_105.20KF.001\_XShooter &  & Jacobs et al. 2019a,b & 37.37911 & -29.13786 \\
AGEL023211+001339A & 2.366 & 3.0 & JONES\_2020B\_U044\_NIRES & 0.893 & 2.0 & JONES\_2020B\_U044\_ESI & 17307, Tran & Jacobs et al. 2019a,b & 38.04683 & 0.22756 \\
AGEL024303-000600A & 1.729 & 3.0 & JONES\_2020B\_U044\_NIRES & 0.367\^ &  & SDSS & 16773, Glazebrook & Jacobs et al. 2019a,b & 40.76267 & -0.10005 \\
AGEL024303-000600B & 2.806 & 2.0 & KACPRZAK\_2019B\_W226\_ESI & 0.367\^ &  & SDSS & 16773, Glazebrook & Jacobs et al. 2019a,b & 40.76267 & -0.10005 \\
AGEL024605-060739A & 2.730 & 3.0 & KACPRZAK\_2021B\_W242\_NIRES &  &  &  &  & Huang et al. 2020 & 41.52050 & -6.12750 \\
AGEL025052-552412A & 2.478 & 3.0 & LOPEZ\_105.20KF.001\_XShooter & 0.722 & 3.0 & LOPEZ\_105.20KF.001\_XShooter &  & Jacobs et al. 2019a,b & 42.71781 & -55.40325 \\
AGEL025220-473238A &  &  &  & 0.494 & 3.0 & GLAZEBROOK\_0101.A-0577\_XShooter & 16773, Glazebrook & Jacobs et al. 2019a,b & 43.08284 & -47.54382 \\
AGEL033203-132510A & 2.335 & 3.0 & KACPRZAK\_2022B\_W211\_NIRES &  &  &  & 16773, Glazebrook & Jacobs et al. 2019a,b & 53.01064 & -13.41950 \\
AGEL033203-132510B & 1.433 & 2.0 & KACPRZAK\_2022B\_W211\_NIRES &  &  &  & 16773, Glazebrook & Jacobs et al. 2019a,b & 53.01064 & -13.41950 \\
AGEL033717-315214A & 1.955 & 3.0 & GLAZEBROOK\_0101.A-0577\_XShooter & 0.525 & 3.0 & GLAZEBROOK\_0101.A-0577\_XShooter &  & Jacobs et al. 2019a,b & 54.32183 & -31.87043 \\
AGEL034131-513045A & 2.035 & 2.0 & LOPEZ\_108.22JL.001\_XShooter &  &  &  & 16773, Glazebrook & Jacobs et al. 2019a,b & 55.37833 & -51.51241 \\
AGEL035346-170639B & 1.460 & 2.0 & JONES\_2021B\_U013\_KCWI & 0.617 & 2.0 & JONES\_2021B\_U013\_KCWI & 16773, Glazebrook & Jacobs et al. 2019a,b & 58.44268 & -17.11090 \\
AGEL035346-170639A & 1.675 & 3.0 & JONES\_2021B\_U013\_KCWI & 0.617 & 2.0 & JONES\_2021B\_U013\_KCWI & 16773, Glazebrook & Jacobs et al. 2019a,b & 58.44268 & -17.11090 \\
AGEL035418-160952A & 1.910 & 1.0 & KACPRZAK\_2020B\_W127\_NIRES & 0.574 & 3.0 & JONES\_2020B\_U044\_ESI &  & Jacobs et al. 2019a,b & 58.57614 & -16.16450 \\
AGEL040823-532714A &  &  &  & 0.641 & 1.0 & GLAZEBROOK\_0101.A-0577\_XShooter & 16773, Glazebrook & Diehl et al. 2017 & 62.09439 & -53.45394 \\
AGEL042439-331742A & 1.188 & 3.0 & GLAZEBROOK\_0101.A-0577\_XShooter & 0.564 & 3.0 & GLAZEBROOK\_0101.A-0577\_XShooter & 16773, Glazebrook & Jacobs et al. 2019a,b & 66.16119 & -33.29491 \\
AGEL043806-322852A & 0.920 & 2.0 & KACPRZAK\_2021B\_W242\_NIRES & 0.343 & 3.0 & LOPEZ\_105.20KF.001\_XShooter & 16773, Glazebrook & Jacobs et al. 2019a,b & 69.52573 & -32.48116 \\
AGEL053724-464702A & 2.344 & 2.0 & LOPEZ\_108.22JL.001\_XShooter & 0.354 & 3.0 & LOPEZ\_108.22JL.001\_XShooter & 16773, Glazebrook & Jacobs et al. 2019a,b & 84.35163 & -46.78401 \\
AGEL061815+501821A & 3.336 & 3.0 & JONES\_2024A\_U039\_KCWI & 0.521 & 3.0 & JONES\_2020B\_U044\_ESI & 15867, Huang & Huang et al. 2021 & 94.56390 & 50.30590 \\
AGEL075524+344540A & 2.635 & 2.0 & JONES\_2020B\_U044\_NIRES & 0.722\^ &  & SDSS & 15867, Huang & Huang et al. 2021 & 118.84800 & 34.76100 \\
AGEL080820+103142B & 1.452 & 3.0 & JONES\_2020B\_U044\_ESI & 0.475\^ &  & SDSS & 15867, Huang & Huang et al. 2020 & 122.08520 & 10.52840 \\
AGEL080820+103142A & 1.303 & 3.0 & JONES\_2020B\_U044\_ESI & 0.475\^ &  & SDSS & 15867, Huang & Huang et al. 2020 & 122.08520 & 10.52840 \\
AGEL083930+021025A & 1.769 & 3.0 & KACPRZAK\_2022B\_W211\_NIRES &  &  &  &  & Jacobs et al. 2019a,b & 129.87662 & 2.17349 \\
AGEL084633-015417A & 2.635 & 2.0 & KACPRZAK\_2022B\_W211\_NIRES &  &  &  &  & Jacobs et al. 2019a,b & 131.63619 & -1.90466 \\
AGEL084943+294328A & 2.057 & 3.0 & KACPRZAK\_2022B\_W211\_NIRES & 0.680\^ &  & SDSS &  & Jacobs et al. 2019a,b & 132.42941 & 29.72437 \\
AGEL085331+232155A & 2.190 & 3.0 & KACPRZAK\_2021A\_W235\_NIRES & 0.306 &  & JONES\_2021A\_U022\_ESI & 15867, Huang & Huang et al. 2021 & 133.38000 & 23.36520 \\
AGEL085413-042409A & 2.393 & 3.0 & NANAYAKKARA\_2022A\_W226\_KCWI &  &  &  & 16773, Glazebrook & Jacobs et al. 2019a,b & 133.55308 & -4.40257 \\
AGEL085917+061517A & 2.658 & 3.0 & KACPRZAK\_2022B\_W211\_NIRES &  &  &  &  & Jacobs et al. 2019a,b & 134.82065 & 6.25486 \\
AGEL090115+095624A & 2.637 & 3.0 & KACPRZAK\_2021B\_W242\_NIRES &  &  &  & 16773, Glazebrook & Huang et al. 2020 & 135.31250 & 9.94010 \\
AGEL091126+141757A & 1.206 & 3.0 & JONES\_2021B\_U012\_ESI & 0.546\^ &  & SDSS &  & Huang et al. 2020 & 137.85680 & 14.29910 \\
AGEL091905+033639A & 2.193 & 3.0 & KACPRZAK\_2022B\_W211\_NIRES & 0.444\^ &  & SDSS &  & Jacobs et al. 2019a,b & 139.76916 & 3.61072 \\
AGEL091935+303156A & 1.812 & 3.0 & KACPRZAK\_2021A\_W235\_NIRES & 0.427\^ &  & SDSS & 17307, Tran & Huang et al. 2021 & 139.89600 & 30.53230 \\
\hline
\end{tabular}
}\caption{Redshifts Measured as part of the AGEL Survey.(1) AGEL ID comprised of RA$\pm$DEC and the arc ID (A for main arc, B for second arc), (2) Source redshift, (3) Source redshift quality flag (3 high, 1 low), (4) Proposal ID for source redshift, (5) Deflector redshift, where a caret (\string^) indicates a literature measurement, (6) Deflector redshift quality flag (3 high, 1 low), (7) Proposal ID for source redshift, (8) Proposal ID and PI of HST photometry, (9) Origin of literature redshift measurement. (10) Right Ascension (11) Declination.\label{tbl:redshift_summary}}
\end{table*}

\begin{table*}
\resizebox{\textwidth}{!}{
\begin{tabular}{c|ccc|ccc|cccc}
\hline
Object Name & $z_{\rm{source}}$ & Source flag & Source Proposal ID & $z_{\rm{deflector}}$ & Deflector flag & Deflector Proposal ID & HST Imaging & Candidate Catalogue & RA & Dec \\
(1) & (2) & (3) & (4) & (5) & (6) & (7) & (8) & (9) & (10) & (11)\\\hline\hline
AGEL092315+182943A & 2.418 & 2.0 & KACPRZAK\_2021A\_W235\_NIRES & 0.873\^ &  & SDSS & 15867, Huang & Huang et al. 2020 & 140.81100 & 18.49540 \\
AGEL093333+091919B & 2.435 & 3.0 & KACPRZAK\_2021A\_W235\_NIRES & 0.743\^ &  & SDSS &  & Jacobs et al. 2019a,b & 143.38871 & 9.32193 \\
AGEL093333+091919A & 2.435 & 3.0 & KACPRZAK\_2021A\_W235\_NIRES & 0.743\^ &  & SDSS &  & Jacobs et al. 2019a,b & 143.38871 & 9.32193 \\
AGEL094328-015453A & 2.125 & 3.0 & KACPRZAK\_2022B\_W211\_NIRES & 0.450\^ &  & SDSS &  & Jacobs et al. 2019a,b & 145.86522 & -1.91482 \\
AGEL094412+322039A & 2.828 & 3.0 & KACPRZAK\_2021A\_W235\_NIRES & 0.595\^ &  & SDSS & 16773, Glazebrook & Jacobs et al. 2019a,b & 146.04930 & 32.34410 \\
AGEL101807-000812A & 1.740 & 2.0 & KACPRZAK\_2021B\_W242\_NIRES & 0.372\^ &  & SDSS & 15867, Huang & Huang et al. 2020 & 154.53070 & -0.13680 \\
AGEL101847-012132A & 1.432 & 3.0 & JONES\_2021A\_U022\_ESI & 0.388\^ &  & SDSS & 15867, Huang & Jacobs et al. 2019a,b & 154.69720 & -1.35900 \\
AGEL103027-064109A & 1.583 & 3.0 & KACPRZAK\_2021A\_W235\_NIRES & 0.468 & 3.0 & JONES\_2021A\_U022\_ESI &  & Jacobs et al. 2019a,b & 157.61347 & -6.68582 \\
AGEL104041+185052A & 0.879 & 3.0 & JONES\_2021A\_U022\_ESI & 0.314\^ &  & SDSS & 17307, Tran & Huang et al. 2021 & 160.17160 & 18.84770 \\
AGEL104056-010359A & 1.210 & 3.0 & KACPRZAK\_2021B\_W242\_NIRES & 0.250\^ &  & SDSS & 15867, Huang & Huang et al. 2020 & 160.23510 & -1.06630 \\
AGEL105100-055628A & 1.940 & 1.0 & KACPRZAK\_2021B\_W242\_NIRES &  &  &  & 16773, Glazebrook & Wittman et al. 2002 & 162.75076 & -5.94108 \\
AGEL110154-060232A & 1.674 & 3.0 & BARONE\_2023B\_W372\_KCWI & 0.483 & 3.0 & BARONE\_2023B\_W372\_KCWI & 15867, Huang & Huang et al. 2021 & 165.47540 & -6.04230 \\
AGEL110245+121111A & 2.806 & 3.0 & KACPRZAK\_2021B\_W242\_NIRES & 0.925 & 2.0 & JONES\_2022A\_U028\_ESI & 15867, Huang & Huang et al. 2020 & 165.68760 & 12.18640 \\
AGEL113929-021826A & 1.553 & 3.0 & KACPRZAK\_2022B\_W211\_NIRES &  &  &  &  & Jacobs et al. 2019a,b & 174.87270 & -2.30722 \\
AGEL114159+191815A & 3.008 & 3.0 & KACPRZAK\_2022B\_W211\_NIRES & 1.443\^ &  & SDSS &  & Jacobs et al. 2019a,b & 175.49611 & 19.30406 \\
AGEL120535+411044A & 2.658 & 3.0 & JONES\_2021B\_U012\_ESI & 0.661 & 3.0 & JONES\_2021B\_U012\_ESI & 15867, Huang & Huang et al. 2021 & 181.39750 & 41.17900 \\
AGEL123809+150151A & 1.161 & 3.0 & JONES\_2021A\_U022\_ESI & 0.572 & 3.0 & JONES\_2021A\_U022\_ESI & 15867, Huang & Jacobs et al. 2019a,b & 189.53702 & 15.03090 \\
AGEL125146+014256A & 1.630 & 3.0 & KACPRZAK\_2021B\_W242\_NIRES & 0.740 & 3.0 & KACPRZAK\_2021B\_W242\_NIRES & 15867, Huang & Huang et al. 2021 & 192.94280 & 1.71550 \\
AGEL132304+034319A & 1.016 & 3.0 & KIRBY\_2021A\_C259\_MOSFIRE & 0.353\^ &  & SDSS & 16773, Glazebrook & Jacobs et al. 2019a,b & 200.76718 & 3.72207 \\
AGEL133041+044015A & 1.169 & 3.0 & KACPRZAK\_2021A\_W235\_NIRES & 0.336\^ &  & SDSS & 15867, Huang & Huang et al. 2020 & 202.66900 & 4.67070 \\
AGEL133145+513431A & 2.052 & 3.0 & KACPRZAK\_2021A\_W235\_NIRES & 0.289\^ &  & SDSS & 15867, Huang & Huang et al. 2021 & 202.93880 & 51.57530 \\
AGEL134333+415503B & 2.843 & 3.0 & JONES\_2024A\_U039\_KCWI & 0.418\^ &  & Diehl et al. 2009 &  & Stark et al. 2013 & 205.88689 & 41.91763 \\
AGEL140839+253104A & 1.289 & 2.0 & JONES\_2021A\_U022\_ESI & 0.663\^ &  & SDSS & 16773, Glazebrook & Jacobs et al. 2019a,b & 212.16144 & 25.51779 \\
AGEL142104+002219A & 2.208 & 3.0 & KACPRZAK\_2021B\_W242\_NIRES &  &  &  & 15867, Huang & Huang et al. 2020 & 215.26540 & 0.37190 \\
AGEL142719-064515A & 1.513 & 3.0 & KACPRZAK\_2021A\_W235\_NIRES & 0.265 & 3.0 & JONES\_2021A\_U022\_ESI & 16773, Glazebrook & Jacobs et al. 2019a,b & 216.82796 & -6.75413 \\
AGEL144133-005401A & 1.667 & 3.0 & NANAYAKKARA\_2022A\_W223\_MOSFIRE & 0.538 &  & JONES\_2021A\_U022\_ESI & 15867, Huang & Jacobs et al. 2019a,b & 220.38753 & -0.90039 \\
AGEL144149+144121B & 2.340 & 3.0 & NANAYAKKARA\_2022A\_W223\_MOSFIRE & 0.741\^ &  & James et al. 2014 & 15867, Huang & Huang et al. 2021 & 220.45490 & 14.68910 \\
AGEL150137+520830A & 2.416 & 3.0 & NANAYAKKARA\_2022A\_W223\_MOSFIRE &  &  &  & 15867, Huang & Huang et al. 2021 & 225.40500 & 52.14170 \\
AGEL150745+052256A & 2.164 & 3.0 & JONES\_2020A\_U160\_NIRES & 0.595\^ &  & SDSS & 16773, Glazebrook & Jacobs et al. 2019a,b & 226.93811 & 5.38230 \\
AGEL150745+052256B & 2.591 & 3.0 & JONES\_2022A\_U041\_KCWI & 0.595\^ &  & SDSS & 16773, Glazebrook & Jacobs et al. 2019a,b & 226.93811 & 5.38230 \\
AGEL150925+390140A & 1.523 & 3.0 & JONES\_2024A\_U039\_KCWI & 0.682\^ &  & SDSS & 15867, Huang & Huang et al. 2021 & 227.35280 & 39.02790 \\
AGEL152509+422753A & 2.261 & 3.0 & KACPRZAK\_2021A\_W235\_NIRES &  &  &  & 15867, Huang & Huang et al. 2021 & 231.28740 & 42.46460 \\
AGEL152560+084639A & 1.951 & 3.0 & JONES\_2022A\_U028\_ESI & 0.602\^ &  & SDSS &  & Jacobs et al. 2019a,b & 231.49978 & 8.77744 \\
AGEL153755+144324A & 2.477 & 3.0 & NANAYAKKARA\_2022A\_W223\_MOSFIRE &  &  &  & 15867, Huang & Huang et al. 2021 & 234.47830 & 14.72320 \\
AGEL153929+165016A & 1.253 & 2.0 & NANAYAKKARA\_2022A\_W223\_MOSFIRE & 0.410\^ &  & SDSS & 15867, Huang & Huang et al. 2021 & 234.87070 & 16.83790 \\
AGEL155417+044339A & 1.721 & 3.0 & KACPRZAK\_2020B\_W127\_NIRES & 0.778 & 3.0 & JONES\_2022A\_U028\_ESI & 15867, Huang & Jacobs et al. 2019a,b & 238.56911 & 4.72756 \\
AGEL162300+213721A & 1.728 & 3.0 & KACPRZAK\_2020B\_W127\_NIRES & 0.758\^ &  & SDSS & 15867, Huang & Jacobs et al. 2019a,b & 245.75139 & 21.62261 \\
AGEL162401+012901A & 2.367 & 3.0 & NANAYAKKARA\_2022A\_W223\_MOSFIRE & 1.092 & 3.0 & JONES\_2022A\_U028\_ESI & 15867, Huang & Huang et al. 2021 & 246.00620 & 1.48360 \\
AGEL165140+280517A & 2.378 & 3.0 & NANAYAKKARA\_2022A\_W223\_MOSFIRE &  &  &  & 15867, Huang & Huang et al. 2021 & 252.91730 & 28.08810 \\
AGEL165742+344858A & 2.465 & 3.0 & JONES\_2020B\_U044\_NIRES &  &  &  & 15867, Huang & Huang et al. 2021 & 254.42350 & 34.81620 \\
AGEL170944+315417A & 2.119 & 2.0 & NANAYAKKARA\_2022A\_W223\_MOSFIRE &  &  &  & 15867, Huang & Huang et al. 2021 & 257.43480 & 31.90460 \\
AGEL171922+244117A & 2.279 & 3.0 & KACPRZAK\_2020B\_W127\_NIRES & 0.529\^ &  & SDSS & 15867, Huang & Jacobs et al. 2019a,b & 259.83959 & 24.68796 \\
AGEL172703+110008A & 1.337 & 3.0 & JONES\_2020B\_U044\_NIRES &  &  &  & 16773, Glazebrook & Jacobs et al. 2019a,b & 261.76369 & 11.00211 \\
AGEL183520+460627A & 3.392 & 3.0 & KACPRZAK\_2020B\_W127\_NIRES &  &  &  & 15867, Huang & Huang et al. 2021 & 278.83380 & 46.10760 \\
AGEL193558+580909A & 3.551 & 3.0 & KACPRZAK\_2020B\_W127\_NIRES & 0.577 & 3.0 & JONES\_2020B\_U044\_ESI & 15867, Huang & Huang et al. 2021 & 293.99270 & 58.15250 \\
AGEL201419-575701A & 2.191 & 3.0 & LOPEZ\_105.20KF.001\_XShooter & 0.717 & 3.0 & LOPEZ\_105.20KF.001\_XShooter & 16773, Glazebrook & Jacobs et al. 2019a,b & 303.58076 & -57.95041 \\
AGEL204312-060954A & 1.892 & 3.0 & KACPRZAK\_2020B\_W127\_NIRES & 0.791 & 3.0 & JONES\_2020B\_U044\_ESI & 15867, Huang & Huang et al. 2021 & 310.80200 & -6.16490 \\
AGEL205616-423857A &  &  &  & 0.720 & 1.0 & LOPEZ\_105.20KF.001\_XShooter & 16773, Glazebrook & Jacobs et al. 2019a,b & 314.06590 & -42.64923 \\
AGEL211005-563931A & 1.182 & 3.0 & LOPEZ\_105.20KF.001\_XShooter & 0.500 & 3.0 & LOPEZ\_105.20KF.001\_XShooter & 16773, Glazebrook & Jacobs et al. 2019a,b & 317.52247 & -56.65849 \\
AGEL211243+000921A & 2.360 & 1.0 & KACPRZAK\_2022B\_W211\_NIRES & 0.445\^ &  & SDSS & 16773, Glazebrook & Jacobs et al. 2019a,b & 318.17974 & 0.15577 \\
AGEL211627-594702A & 1.412 & 3.0 & GLAZEBROOK\_0101.A-0577\_XShooter & 0.395 & 1.0 & GLAZEBROOK\_0101.A-0577\_XShooter & 16773, Glazebrook & Jacobs et al. 2019a,b & 319.11382 & -59.78382 \\
AGEL212252-005949A & 0.928 & 3.0 & JELTEMA\_2024B\_U225\_KCWI & 0.349 & 3.0 & JELTEMA\_2024B\_U225\_KCWI & 17307, Tran & Diehl et al. 2017 & 320.71667 & -0.99700 \\
AGEL212326+015312A & 1.180 & 3.0 & JONES\_2019B\_U058\_ESI & 0.591\^ &  & SDSS & 16773, Glazebrook & Huang et al. 2021 & 320.85840 & 1.88670 \\
AGEL212512-650427A & 2.223 & 2.0 & GLAZEBROOK\_0101.A-0577\_XShooter & 0.779 & 3.0 & GLAZEBROOK\_0101.A-0577\_XShooter & 16773, Glazebrook & Jacobs et al. 2019a,b & 321.30012 & -65.07408 \\
AGEL213758-012924A & 1.458 & 3.0 & KACPRZAK\_2019B\_W226\_ESI & 0.273 & 3.0 & KACPRZAK\_2019B\_W226\_ESI & 16773, Glazebrook & Jacobs et al. 2019a,b & 324.49176 & -1.48996 \\
AGEL214915-001252A & 1.945 & 3.0 & JONES\_2020B\_U044\_NIRES & 0.453\^ &  & SDSS & 16773, Glazebrook & Jacobs et al. 2019a,b & 327.31380 & -0.21431 \\
AGEL215122+134718A & 0.890 & 2.0 & JONES\_2020B\_U044\_ESI & 0.206 & 3.0 & JONES\_2020B\_U044\_ESI & 15867, Huang & Huang et al. 2021 & 327.84080 & 13.78840 \\
AGEL215844+025730A & 2.081 & 3.0 & KACPRZAK\_2020B\_W127\_NIRES & 0.287\^ &  & SDSS & 15867, Huang & Huang et al. 2021 & 329.68200 & 2.95840 \\
AGEL221912-434835A & 2.168 & 3.0 & GLAZEBROOK\_0101.A-0577\_XShooter & 0.710 & 1.0 & GLAZEBROOK\_0101.A-0577\_XShooter & 16773, Glazebrook & Jacobs et al. 2019a,b & 334.80166 & -43.80975 \\
AGEL222609+004142A & 1.896 & 3.0 & JONES\_2020B\_U044\_NIRES & 0.647\^ &  & SDSS & 16773, Glazebrook & Jacobs et al. 2019a,b & 336.53876 & 0.69504 \\
AGEL224405+275916A & 0.960 & 3.0 & JONES\_2019B\_U058\_ESI & 0.343\^ &  & SDSS & 16773, Glazebrook & Huang et al. 2021 & 341.02060 & 27.98770 \\
AGEL224621+223338A & 2.257 & 3.0 & JONES\_2019B\_U058\_ESI &  &  &  &  & Stark et al. 2013 & 341.58817 & 22.56045 \\
AGEL230305-511502A & 2.568 & 3.0 & LOPEZ\_105.20KF.001\_XShooter & 0.514 & 3.0 & LOPEZ\_105.20KF.001\_XShooter & 16773, Glazebrook & Jacobs et al. 2019a,b & 345.76962 & -51.25050 \\
AGEL230522-000212A & 1.838 & 3.0 & KACPRZAK\_2019B\_W226\_ESI & 0.492\^ &  & SDSS &  & Jacobs et al. 2019a,b & 346.34025 & -0.03655 \\
AGEL231112-454658A & 1.555 & 3.0 & LOPEZ\_105.20KF.001\_XShooter & 0.504 & 3.0 & LOPEZ\_105.20KF.001\_XShooter &  & Jacobs et al. 2019a,b & 347.79872 & -45.78281 \\
AGEL231812-110604A & 2.656 & 2.0 & KACPRZAK\_2022B\_W211\_NIRES &  &  &  & 15867, Huang & Huang et al. 2020 & 349.54920 & -11.10120 \\
AGEL231935+115016A & 1.990 & 3.0 & JONES\_2019B\_U058\_ESI & 0.542 & 3.0 & JONES\_2019B\_U058\_ESI &  & Jacobs et al. 2019a,b & 349.89385 & 11.83776 \\
AGEL232128-463049A & 1.749 & 2.0 & LOPEZ\_108.22JL.001\_XShooter &  &  &  & 16773, Glazebrook & Jacobs et al. 2019a,b & 350.36821 & -46.51371 \\
AGEL233459-640407A & 2.495 & 3.0 & LOPEZ\_105.20KF.001\_XShooter & 0.701 & 2.0 & LOPEZ\_105.20KF.001\_XShooter & 16773, Glazebrook & Jacobs et al. 2019a,b & 353.74665 & -64.06860 \\
AGEL233552-515218A & 2.225 & 3.0 & GLAZEBROOK\_0101.A-0577\_XShooter & 0.566 & 3.0 & GLAZEBROOK\_0101.A-0577\_XShooter & 16773, Glazebrook & Jacobs et al. 2019a,b & 353.96636 & -51.87161 \\
AGEL233610-020735A & 2.663 & 3.0 & JONES\_2020B\_U044\_NIRES & 0.494\^ &  & SDSS & 16773, Glazebrook & Huang et al. 2021 & 354.04280 & -2.12640 \\
AGEL234930-511339A & 1.393 & 2.0 & LOPEZ\_108.22JL.001\_XShooter &  &  &  & 16773, Glazebrook & Jacobs et al. 2019a,b & 357.37523 & -51.22751 \\
AGEL235934+020824A & 1.119 & 3.0 & JELTEMA\_2023B\_U039\_NIRES & 0.430 & 3.0 & JELTEMA\_2022B\_U099\_DEIMOS &  & Huang et al. 2020 & 359.88970 & 2.13990 \\
\hline
\end{tabular}
}\caption{Table \ref{tbl:redshift_summary} continued.\label{tbl:redshift_summary_extended}}
\end{table*}

\begin{acknowledgments}
Parts of this research were conducted by the Australian Research Council Centre of Excellence for All Sky Astrophysics in 3 Dimensions (ASTRO 3D), through project number CE170100013.
T.M.B, K.G, D.J.B, and G.F.L acknowledge support from Australian Research Council Discovery Project grant DP230101775.
TJ, KVGC, and FD gratefully acknowledge support from the National Science Foundation through grant AST-2108515, the Gordon and Betty Moore Foundation through Grant GBMF8549, and a UC Davis Chancellor's Fellowship.
X.H. acknowledges the University of San Francisco Faculty Development Fund. Support for HST program 15867 was provided by NASA through a grant from the Space Telescope Science Institute, which is operated by the Association of Universities for Research in Astronomy, Inc., under NASA contract NAS 5-26555.
S.L. and N.T. acknowledge support by FONDECYT grant 1231187.
S.M.S acknowledges funding from the Australian Research Council (DE220100003).
Some of the data presented herein were obtained at the W. M. Keck Observatory, which is operated as a scientific partnership among the California Institute of Technology, the University of California and the National Aeronautics and Space Administration. The Observatory was made possible by the generous financial support of the W. M. Keck Foundation.
The authors wish to recognize and acknowledge the very significant cultural role and reverence that the summit of Maunakea has always had within the indigenous Hawaiian community. We are most fortunate to have the opportunity to conduct observations from this mountain.

\end{acknowledgments}

\facilities{HST(WFC3), HST(ACS), Keck(NIRES), Keck(ESI), Keck(LRIS), Keck(KCWI), Keck(DEIMOS), Keck(MOSFIRE), VLT(X-shooter)}

\software{Astropy, NumPy, Matplotlib, Montage, Pandas, reproject.}

\clearpage
\appendix

\setcounter{table}{0}
\renewcommand{\thetable}{A\arabic{table}}

\begin{table*}
\centering\resizebox*{!}{\textheight}{
\begin{tabular}{ccccc}
\hline
Position & Object Name & HST Name & RA & Dec \\
(1) & (2) & (3) & (4) & (5) \\\hline\hline
\textcolor{green}{1} & AGEL000316-334804A & DESJ0003-3348 & 0.81830 & -33.80120 \\
\textcolor{green}{2} & AGEL000645-442950A & DESJ0006-4429 & 1.68590 & -44.49740 \\
\textcolor{green}{3} & AGEL001310+004004A & DESJ0013+0040 & 3.29020 & 0.66770 \\
\textcolor{green}{4} & AGEL004257-371858A & DESJ0042-3718 & 10.73880 & -37.31620 \\
\textcolor{green}{5} & AGEL010238+015857A & DESJ0102+0158 & 15.65960 & 1.98240 \\
\textcolor{green}{6} & AGEL013355-643413A & DESJ0133-6434 & 23.47770 & -64.57030 \\
\textcolor{green}{7} & AGEL014106-171324A & DESJ0141-1713 & 25.27560 & -17.22330 \\
\textcolor{green}{8} & AGEL014253-183116A & DESJ0142-1831 & 25.72030 & -18.52110 \\
\textcolor{green}{9} & AGEL020613-011417A & DESJ0206-0114 & 31.55610 & -1.23820 \\
\textcolor{green}{10} & AGEL021225-085211A & DESJ0212-0852 & 33.10510 & -8.86970 \\
\textcolor{green}{11} & AGEL024303-000600A & DESJ0243-0006 & 40.76270 & -0.10010 \\
\textcolor{green}{12} & AGEL035346-170639A & DCLS0353-1706 & 58.44270 & -17.11090 \\
\textcolor{green}{13} & AGEL042439-331742A & DESJ0424-3317 & 66.16120 & -33.29490 \\
\textcolor{green}{14} & AGEL043806-322852A & DESJ0438-3228 & 69.52570 & -32.48120 \\
\textcolor{green}{15} & AGEL053724-464702A & DESJ0537-4647 & 84.35160 & -46.78400 \\
\textcolor{green}{16} & AGEL094412+322039A & DCLS0944+3220 & 146.04930 & 32.34410 \\
\textcolor{green}{17} & AGEL132304+034319A & DCLS1323+0343 & 200.76720 & 3.72210 \\
\textcolor{green}{18} & AGEL140839+253104A & DCLS1408+2531 & 212.16140 & 25.51780 \\
\textcolor{green}{19} & AGEL142719-064515A & DCLS1427-0645 & 216.82800 & -6.75410 \\
\textcolor{green}{20} & AGEL150745+052256A & DCLS1507+0522 & 226.93810 & 5.38230 \\
\textcolor{green}{21} & AGEL201419-575701A & DESJ2014-5757 & 303.58080 & -57.95040 \\
\textcolor{green}{22} & AGEL211005-563931A & DESJ2110-5639 & 317.52250 & -56.65850 \\
\textcolor{green}{23} & AGEL211243+000921A & DESJ2112+0009 & 318.17970 & 0.15580 \\
\textcolor{green}{24} & AGEL211627-594702A & DESJ2116-5947 & 319.11380 & -59.78380 \\
\textcolor{green}{25} & AGEL212326+015312A & CSWA157 & 320.85840 & 1.88670 \\
\textcolor{green}{26} & AGEL212512-650427A & DESJ2125-6504 & 321.30010 & -65.07410 \\
\textcolor{green}{27} & AGEL213758-012924A & DESJ2137-0129 & 324.49180 & -1.49000 \\
\textcolor{green}{28} & AGEL214915-001252A & DESJ2149-0012 & 327.31380 & -0.21430 \\
\textcolor{green}{29} & AGEL221912-434835A & DESJ2219-4348 & 334.80170 & -43.80980 \\
\textcolor{green}{30} & AGEL222609+004142A & DESJ2226+0041 & 336.53880 & 0.69500 \\
\textcolor{green}{31} & AGEL224405+275916A & CSWA129 & 341.02060 & 27.98770 \\
\textcolor{green}{32} & AGEL230305-511502A & DESJ2303-5115 & 345.76960 & -51.25050 \\
\textcolor{green}{33} & AGEL233459-640407A & DESJ2334-6404 & 353.74670 & -64.06860 \\
\textcolor{green}{34} & AGEL233552-515218A & DESJ2335-5152 & 353.96640 & -51.87160 \\
\textcolor{green}{35} & AGEL233610-020735A & PS1J2336 & 354.04280 & -2.12640 \\
\textcolor{YellowOrange}{36} & AGEL000729-443446A & DESJ0007-4434 & 1.87200 & -44.57950 \\
\textcolor{YellowOrange}{37} & AGEL010158-491738A & DESJ0101-4917 & 15.49180 & -49.29390 \\
\textcolor{YellowOrange}{38} & AGEL012429-291856A & DESJ0124-2918 & 21.11890 & -29.31560 \\
\textcolor{YellowOrange}{39} & AGEL022709-471856A & DESJ0227-4718 & 36.78730 & -47.31550 \\
\textcolor{YellowOrange}{40} & AGEL025220-473238A & DESJ0252-4732 & 43.08280 & -47.54380 \\
\textcolor{YellowOrange}{41} & AGEL033203-132510A & DCLS0332-1325 & 53.01060 & -13.41950 \\
\textcolor{YellowOrange}{42} & AGEL034131-513045A & DESJ0341-5130 & 55.37830 & -51.51240 \\
\textcolor{YellowOrange}{43} & AGEL040823-532714A & DESJ0408-5327 & 62.09440 & -53.45390 \\
\textcolor{YellowOrange}{44} & AGEL085413-042409A & DCLS0854-0424 & 133.55310 & -4.40260 \\
\textcolor{YellowOrange}{45} & AGEL090115+095624A & DESI-135.3125+09.9401 & 135.31250 & 9.94010 \\
\textcolor{YellowOrange}{46} & AGEL105100-055628A & DLS432021848 & 162.75080 & -5.94110 \\
\textcolor{YellowOrange}{47} & AGEL172703+110008A & DCLS1727+1100 & 261.76370 & 11.00210 \\
\textcolor{YellowOrange}{48} & AGEL205616-423857A & DESJ2056-4238 & 314.06590 & -42.64920 \\
\textcolor{YellowOrange}{49} & AGEL232128-463049A & DESJ2321-4630 & 350.36820 & -46.51370 \\
\textcolor{YellowOrange}{50} & AGEL234930-511339A & DESJ2349-5113 & 357.37520 & -51.22750 \\
\textcolor{CadetBlue}{51} & AGEL001030-431515A & DESJ0010-4315 & 2.62680 & -43.25410 \\
\textcolor{CadetBlue}{52} & AGEL010520+014457A & DESJ0105+0144 & 16.33190 & 1.74900 \\
\textcolor{CadetBlue}{53} & AGEL021408-020629A & DCLS0214-0206 & 33.53340 & -2.10790 \\
\textcolor{CadetBlue}{54} & AGEL024229-294305A & DESJ0242-2943 & 40.62050 & -29.71820 \\
\textcolor{CadetBlue}{55} & AGEL025029-410418A & DESJ0250-4104 & 42.62080 & -41.07170 \\
\textcolor{CadetBlue}{56} & AGEL030022-500129A & DESJ0300-5001 & 45.09020 & -50.02470 \\
\textcolor{CadetBlue}{57} & AGEL032216-523440A & DESJ0322-5234 & 50.56840 & -52.57790 \\
\textcolor{CadetBlue}{58} & AGEL032904-565658A & DESJ0329-5656 & 52.26620 & -56.94940 \\
\textcolor{CadetBlue}{59} & AGEL035606-560729A & DESJ0356-5607 & 59.02600 & -56.12480 \\
\textcolor{CadetBlue}{60} & AGEL041645-552500A & DESJ0416-5525 & 64.18670 & -55.41680 \\
\textcolor{CadetBlue}{61} & AGEL042816-321800A & DESJ0428-3218 & 67.06770 & -32.30000 \\
\textcolor{CadetBlue}{62} & AGEL053746-471120A & DESJ0537-4711 & 84.44090 & -47.18900 \\
\textcolor{CadetBlue}{63} & AGEL060357-355806A & DESJ0603-3558 & 90.98540 & -35.96830 \\
\textcolor{CadetBlue}{64} & AGEL132227-050135A & DCLS1322-0501 & 200.61330 & -5.02630 \\
\textcolor{CadetBlue}{65} & AGEL144431+241843A & DCLS1444+2418 & 221.13010 & 24.31190 \\
\textcolor{CadetBlue}{66} & AGEL144640-000350A & DCLS1446-0003 & 221.66720 & -0.06400 \\
\textcolor{CadetBlue}{67} & AGEL144743-065709A & DCLS1447-0657 & 221.92780 & -6.95260 \\
\textcolor{CadetBlue}{68} & AGEL144802-081249A & DCLS1448-0812 & 222.00790 & -8.21370 \\
\textcolor{CadetBlue}{69} & AGEL203911-545945A & DESJ2039-5459 & 309.79780 & -54.99590 \\
\textcolor{CadetBlue}{70} & AGEL204555+003148A & DCLS2045+0031 & 311.47710 & 0.53000 \\
\textcolor{CadetBlue}{71} & AGEL212447-412816A & DESJ2124-4128 & 321.19660 & -41.47100 \\
\hline
\end{tabular}
}\caption{Object names for targets in Figure \ref{FIG:HST_SNAP_MEGAFIG} (ID:16773, PI Glazebrook). (1) The subpanel position, counting from left to right, top to bottom. The colour reflects whether the object has both source and deflector redshifts (green), one of source or deflector redshift (orange), or no available redshifts (grey). (2) The AGEL object name. (3) The object name used in the HST observations. (4), (5) Right ascension and declination in decimal degrees.\label{tbl:hst_snap_megafig}}
\end{table*}

\begin{table*}
\centering\resizebox*{!}{\textheight}{
\begin{tabular}{ccccc}
\hline
Position & Object Name & HST Name & RA & Dec \\
(1) & (2) & (3) & (4) & (5) \\\hline\hline
\textcolor{green}{1} & AGEL212252-005949A & DES2122-0059 & 320.71667 & -0.99700 \\
\textcolor{green}{2} & AGEL091935+303156A & DLS210207 & 139.89600 & 30.53230 \\
\textcolor{green}{3} & AGEL104041+185052A & DESI-160 & 160.17160 & 18.84770 \\
\textcolor{green}{4} & AGEL014327-085021A & DESJ0143-0850 & 25.86222 & -8.83925 \\
\textcolor{green}{5} & AGEL023211+001339A & HSCJ023211+001339 & 38.04683 & 0.22756 \\
\textcolor{green}{6} & AGEL011759-052718A & DESJ0117-0527 & 19.49477 & -5.45492 \\
\textcolor{green}{7} & AGEL014504-045551A & DESJ0145-0455 & 26.26791 & -4.93084 \\
\textcolor{YellowOrange}{8} & AGEL013003-374458A & DES0130-3744 & 22.51201 & -37.74938 \\
\textcolor{CadetBlue}{9} & AGEL053349-253654A & DESJ0533-2536 & 83.45553 & -25.61511 \\
\textcolor{CadetBlue}{10} & AGEL103255+751854A & DESI-158.2306+75.3149 & 158.23060 & 75.31490 \\
\textcolor{CadetBlue}{11} & AGEL221638-441920A & DESJ2216-4419 & 334.15922 & -44.32216 \\
\textcolor{CadetBlue}{12} & AGEL091433+301620A & DCLS0914+3016 & 138.63602 & 30.27227 \\
\textcolor{CadetBlue}{13} & AGEL085920+391702A & DESI-134.8336+39.2838 & 134.83363 & 39.28376 \\
\textcolor{CadetBlue}{14} & AGEL101103-001649A & DELJ101103-001649 & 152.76540 & -0.28050 \\
\textcolor{CadetBlue}{15} & AGEL001030-431515A & DESJ0010-4315 & 2.62678 & -43.25413 \\
\textcolor{CadetBlue}{16} & AGEL102643-023319A & DELJ102643-023319 & 156.67940 & -2.55540 \\
\textcolor{CadetBlue}{17} & AGEL103858-354726A & DELJ103858-354726 & 159.74540 & -35.79070 \\
\textcolor{CadetBlue}{18} & AGEL111800-153227A & DELJ111800-153227 & 169.50380 & -15.54100 \\
\textcolor{CadetBlue}{19} & AGEL151048+630325A & DESI-227.7011+63.0570 & 227.70111 & 63.05703 \\
\textcolor{CadetBlue}{20} & AGEL082017+643112A & DESI-125.0716+64.5201 & 125.07158 & 64.52011 \\
\textcolor{CadetBlue}{21} & AGEL113521+714307A & DESI-173.8651+71.7367 & 173.86513 & 71.73673 \\
\textcolor{CadetBlue}{22} & AGEL224504-501725A & DESJ2245-5017 & 341.26572 & -50.29037 \\
\textcolor{CadetBlue}{23} & AGEL212707-514951A & DESJ2127-5149 & 321.77933 & -51.83083 \\
\textcolor{CadetBlue}{24} & AGEL110419-425746A & DELJ110419-425746 & 166.08090 & -42.96280 \\
\textcolor{CadetBlue}{25} & AGEL012259-022705A & DES0122-0207 & 20.74568 & -2.45147 \\
\textcolor{CadetBlue}{26} & AGEL225607+012534A & HSCJ225607+012534 & 344.03111 & 1.42638 \\
\textcolor{CadetBlue}{27} & AGEL210607-441154A & DESJ2106-4411 & 316.52989 & -44.19821 \\
\textcolor{CadetBlue}{28} & AGEL165046+612939A & DESI-252.6907+61.4942 & 252.69066 & 61.49416 \\
\textcolor{CadetBlue}{29} & AGEL041810-545735A & DESJ0418-5457 & 64.54117 & -54.95973 \\
\textcolor{CadetBlue}{30} & AGEL112354+505149A & DESI-171.0040+50.8683 & 171.00404 & 50.86830 \\
\textcolor{CadetBlue}{31} & AGEL170332+631417A & DESI-255.8840+63.2381 & 255.88400 & 63.23810 \\
\textcolor{CadetBlue}{32} & AGEL184141+741405A & DESI-280.4220+74.2348 & 280.42200 & 74.23480 \\
\textcolor{CadetBlue}{33} & AGEL015825-003959A & DES0158-0040 & 29.60320 & -0.66649 \\
\textcolor{CadetBlue}{34} & AGEL042225-403156A & DESJ0422-4031 & 65.60553 & -40.53217 \\
\textcolor{CadetBlue}{35} & AGEL035714-475647A & DES0357-4756 & 59.30752 & -47.94633 \\
\textcolor{CadetBlue}{36} & AGEL014359+125610A & DCLS0143+1256 & 25.99491 & 12.93613 \\
\textcolor{CadetBlue}{37} & AGEL030416-492126A & DES0304-4921 & 46.06730 & -49.35724 \\
\textcolor{CadetBlue}{38} & AGEL012441-015734A & DCLS0124-0157 & 21.17193 & -1.95944 \\
\textcolor{CadetBlue}{39} & AGEL073726+665837A & DESI-114.3573+66.9769 & 114.35725 & 66.97694 \\
\textcolor{CadetBlue}{40} & AGEL005055-172033A & DCLS0050-1720 & 12.73022 & -17.34243 \\
\hline
\end{tabular}
}\caption{Object names for targets in \ref{FIG:GAP_HST_FIG} (ID:17307, PI Tran). (1) The subpanel position, counting from left to right, top to bottom. The colour reflects whether the object has both source and deflector redshifts (green), one of source or deflector redshift (orange), or no available redshifts (grey). (2) The AGEL object name. (3) The object name used in the HST observations. (4), (5) Right ascension and declination in decimal degrees.\label{tbl:GAP_hist_fig}}
\end{table*}

\begin{table*}
\centering\resizebox*{!}{\textheight}{
\begin{tabular}{ccccc}
\hline
Position & Object Name & HST Name & RA & Dec \\
(1) & (2) & (3) & (4) & (5) \\\hline\hline
\textcolor{green}{1} & AGEL002527+101107A & DESI-006.3643+10.1853 & 6.36430 & 10.18530 \\
\textcolor{green}{2} & AGEL013442+043350A & DESI-023.6765+04.5639 & 23.67650 & 4.56390 \\
\textcolor{green}{3} & AGEL013639+000818A & DESI-024.1631+00.1384 & 24.16310 & 0.13836 \\
\textcolor{green}{4} & AGEL061815+501821A & DESI-094.5639+50.3059 & 94.56390 & 50.30590 \\
\textcolor{green}{5} & AGEL075524+344540A & DESI-118.8480+34.7610 & 118.84800 & 34.76100 \\
\textcolor{green}{6} & AGEL080820+103142A & DESI-122.0852+10.5284 & 122.08520 & 10.52840 \\
\textcolor{green}{7} & AGEL085331+232155A & DESI-133.3800+23.3652 & 133.38000 & 23.36520 \\
\textcolor{green}{8} & AGEL092315+182943A & DESI-140.8110+18.4954 & 140.81100 & 18.49540 \\
\textcolor{green}{9} & AGEL101807-000812A & DESI-154.5307-00.1368 & 154.53070 & -0.13680 \\
\textcolor{green}{10} & AGEL101847-012132A & DESI-154.6972-01.3590 & 154.69720 & -1.35900 \\
\textcolor{green}{11} & AGEL104056-010359A & DESI-160.2351-01.0663 & 160.23510 & -1.06630 \\
\textcolor{green}{12} & AGEL110154-060232A & DESI-165.4754-06.0423 & 165.47540 & -6.04230 \\
\textcolor{green}{13} & AGEL110245+121111A & DESI-165.6876+12.1864 & 165.68760 & 12.18640 \\
\textcolor{green}{14} & AGEL120535+411044A & DESI-181.3974+41.1790 & 181.39750 & 41.17900 \\
\textcolor{green}{15} & AGEL123809+150151A & DESI-189.5370+15.0309 & 189.53702 & 15.03090 \\
\textcolor{green}{16} & AGEL125146+014256A & DESI-192.9428+01.7155 & 192.94280 & 1.71550 \\
\textcolor{green}{17} & AGEL133041+044015A & DESI-202.6690+04.6707 & 202.66900 & 4.67070 \\
\textcolor{green}{18} & AGEL133145+513431A & DESI-202.9388+51.5753 & 202.93880 & 51.57530 \\
\textcolor{green}{19} & AGEL144133-005401A & DESI-220.3861-00.8995 & 220.38753 & -0.90039 \\
\textcolor{green}{20} & AGEL150925+390140A & DESI-227.3528+39.0279 & 227.35280 & 39.02790 \\
\textcolor{green}{21} & AGEL153929+165016A & DESI-234.8707+16.8379 & 234.87070 & 16.83790 \\
\textcolor{green}{22} & AGEL155417+044339A & DESI-238.5690+04.7276 & 238.56911 & 4.72756 \\
\textcolor{green}{23} & AGEL162300+213721A & DESI-245.7514+21.6226 & 245.75139 & 21.62261 \\
\textcolor{green}{24} & AGEL162401+012901A & DESI-246.0062+01.4836 & 246.00620 & 1.48360 \\
\textcolor{green}{25} & AGEL171922+244117A & DESI-259.8396+24.6880 & 259.83959 & 24.68796 \\
\textcolor{green}{26} & AGEL193558+580909A & DESI-293.9927+58.1525 & 293.99270 & 58.15250 \\
\textcolor{green}{27} & AGEL204312-060954A & DESI-310.8019-06.1649 & 310.80200 & -6.16490 \\
\textcolor{green}{28} & AGEL215122+134718A & DESI-327.8408+13.7884 & 327.84080 & 13.78840 \\
\textcolor{green}{29} & AGEL215844+025730A & DESI-329.6820+02.9584 & 329.68200 & 2.95840 \\
\textcolor{YellowOrange}{30} & AGEL001702-100911A & DESI-004.2564-10.1530 & 4.25640 & -10.15300 \\
\textcolor{YellowOrange}{31} & AGEL142104+002219A & DESI-215.2654+00.3719 & 215.26540 & 0.37190 \\
\textcolor{YellowOrange}{32} & AGEL150137+520830A & DESI-225.4050+52.1417 & 225.40500 & 52.14170 \\
\textcolor{YellowOrange}{33} & AGEL152509+422753A & DESI-231.2858+42.4643 & 231.28740 & 42.46460 \\
\textcolor{YellowOrange}{34} & AGEL153755+144324A & DESI-234.4783+14.7232 & 234.47830 & 14.72320 \\
\textcolor{YellowOrange}{35} & AGEL165140+280517A & DESI-252.9173+28.0881 & 252.91730 & 28.08810 \\
\textcolor{YellowOrange}{36} & AGEL165742+344858A & DESI-254.4235+34.8162 & 254.42350 & 34.81620 \\
\textcolor{YellowOrange}{37} & AGEL170944+315417A & DESI-257.4348+31.9046 & 257.43480 & 31.90460 \\
\textcolor{YellowOrange}{38} & AGEL183520+460627A & DESI-278.8338+46.1076 & 278.83380 & 46.10760 \\
\textcolor{YellowOrange}{39} & AGEL231812-110604A & DESI-349.5492-11.1012 & 349.54920 & -11.10120 \\
\textcolor{CadetBlue}{40} & AGEL013204-160014A & DESI-023.0157-16.0040 & 23.01570 & -16.00400 \\
\textcolor{CadetBlue}{41} & AGEL014156+304531A & DESI-025.4848+30.7585 & 25.48480 & 30.75850 \\
\textcolor{CadetBlue}{42} & AGEL020145-273942A & DESI-030.4360-27.6618 & 30.43611 & -27.66177 \\
\textcolor{CadetBlue}{43} & AGEL021514-290925A & DESI-033.8095-29.1570 & 33.80950 & -29.15700 \\
\textcolor{CadetBlue}{44} & AGEL044821-192502A & DESI-072.0873-19.4172 & 72.08730 & -19.41730 \\
\textcolor{CadetBlue}{45} & AGEL122537-072508A & DESI-186.4028-07.4188 & 186.40280 & -7.41880 \\
\textcolor{CadetBlue}{46} & AGEL122719+172557A & DESI-186.8292+17.4324 & 186.82920 & 17.43240 \\
\textcolor{CadetBlue}{47} & AGEL123736+553343A & DESI-189.4008+55.5619 & 189.40080 & 55.56190 \\
\textcolor{CadetBlue}{48} & AGEL142749+081045A & DESI-216.9538+08.1792 & 216.95385 & 8.17917 \\
\textcolor{CadetBlue}{49} & AGEL144149+144121A & DESI-220.4549+14.6891 & 220.45490 & 14.68910 \\
\textcolor{CadetBlue}{50} & AGEL150511+172042A & DESI-226.2950+17.3451 & 226.29500 & 17.34510 \\
\textcolor{CadetBlue}{51} & AGEL151009+203725A & DESI-227.5364+20.6236 & 227.53640 & 20.62360 \\
\hline
\end{tabular}
}\caption{Object names for targets in \ref{FIG:HUANG_SNAP_FIG} (ID:15867, PI Huang). (1) The subpanel position, counting from left to right, top to bottom. The colour reflects whether the object has both source and deflector redshifts (green), one of source or deflector redshift (orange), or no available redshifts (grey). (2) The AGEL object name. (3) The object name used in the HST observations. (4), (5) Right ascension and declination in decimal degrees.\label{tbl:huang_snap_fig}}
\end{table*}

\bibliographystyle{aasjournal}

\end{document}